\documentclass[twocolumn,showpacs,preprintnumbers,amsmath,amssymb,prb,superscriptaddress]{revtex4-1}

\usepackage{graphicx}
\usepackage{dcolumn}
\usepackage{overpic}
\usepackage{float} 
\usepackage{bbm}
\usepackage{color}

\usepackage{amssymb}
\usepackage{amsmath}

\usepackage{appendix}

\newcommand{\ket}[1]{\left|#1\right>}
\newcommand{\bra}[1]{\left<#1\right|}

\newcommand{\nn}{\nonumber\\}

\newcommand{\f}[1]{\mbox{\boldmath$#1$}}

\newcommand{\bea}{\begin{eqnarray}}
\newcommand{\eea}{\end{eqnarray}}
\newcommand{\beann}{\begin{eqnarray*}}
\newcommand{\eeann}{\end{eqnarray*}}

\newcommand{\trace}[1]{{\rm Tr}\left\{ #1 \right\}}

\newcommand{\abs}[1]{{\left| #1 \right|}}

\newcommand{\ii}{\mathrm{i}}  

\begin{document}

\title{Thermoelectric performance of topological boundary modes} 

\author{S. B\"ohling}
\email{sina.boehling@campus.tu-berlin.de}
\affiliation{Institut f\"ur Theoretische Physik, Technische Universit\"at Berlin, Hardenbergstr. 36, 10623 Berlin, Germany}

\author{G. Engelhardt}
\email{georg.engelhardt@csrc.ac.cn}
\affiliation{Beijing Computational Science Research Center, Beijing 100193, Peopleʼs Republic of China}

\author{G. Platero}
\email{gplatero@icmm.csic.es}
\affiliation{Instituto de Ciencia de Materiales de Madrid, CSIC, 28049 Madrid, Spain}

\author{G. Schaller}
\email{gernot.schaller@tu-berlin.de}
\affiliation{Institut f\"ur Theoretische Physik, Technische Universit\"at Berlin, Hardenbergstr. 36, 10623 Berlin, Germany}

\begin{abstract} 
We investigate quantum transport and thermoelectrical properties of a finite-size Su-Schrieffer-Heeger model, 
a paradigmatic model for a one-dimensional topological insulator, which displays topologically
protected edge states.
By coupling the model to two fermionic reservoirs at its ends, we can explore the non-equilibrium dynamics of the system.
Investigating the energy-resolved transmission, the current and the noise, we find that these observables 
can be used to detect the topologically non-trivial phase. 
With specific parameters and asymmetric reservoir coupling strengths, we show that we can dissipatively prepare the edge states as stationary states of a non-equilibrium configuration.
In addition, we point out that the edge states can be exploited to design a refrigerator driven by chemical work or a heat engine driven by a thermal gradient, 
respectively.
These thermal devices do not require asymmetric couplings and are topologically protected against symmetry-preserving perturbations. Their maximum efficiencies significantly exceed that of a single quantum dot device at comparable coupling strengths.
\end{abstract}

\maketitle


\section{Introduction}

Non-trivial topological band structures appear in various contexts~\cite{Hasan2010,Asboth2016,Chiu2016}. 
Starting from the integer quantum Hall effect and the spin-quantum Hall effect observed in solid-state systems, 
there has been also experimental and theoretical effort to discover topology-related effects in phononic and 
photonic systems~\cite{Peano,Peano2016,Engelhardt2016}, 
cold-atom experiments~\cite{Aidelsburger2013,Jotzu2014} or mechanical systems~\cite{Suesstrunk2015,Lee2017}.

Materials and devices based on topological band structures can give rise to new technical innovations. 
For example, the edge channels of topological insulators allow for a dissipationless energy transfer via scatter-free currents along the edges~\cite{Koenig2007,chang2015a,Klitzing1986,Roth2009,Dzero2016}.
Moreover, Majorana fermions appearing at the edge of a topological superconductor are topologically protected against (symmetry-respecting) perturbations, which
makes them interesting candidates for quantum computation~\cite{Freedman2003,das_sarma2015a,Nayak2008}.
In photonic systems, it has been suggested to use topological bandstructures to create a non-reciprocal chiral circulator~\cite{Peano2016}, 
and topological networks have been proposed as robust information transmitters~\cite{lang2017a}.

\begin{figure}[ht!]
    \centering 	
    \setlength{\unitlength}{.1\linewidth}
	\begin{picture}(10,6.8)
	\put(0.,4.6){\includegraphics[width=\linewidth]{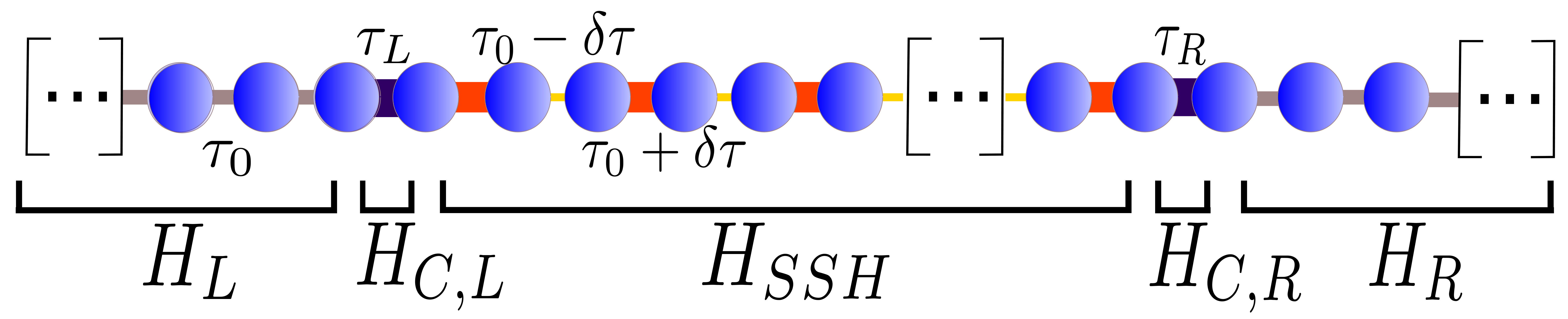}}
	\put(0.,0.){\includegraphics[width=.49\linewidth]{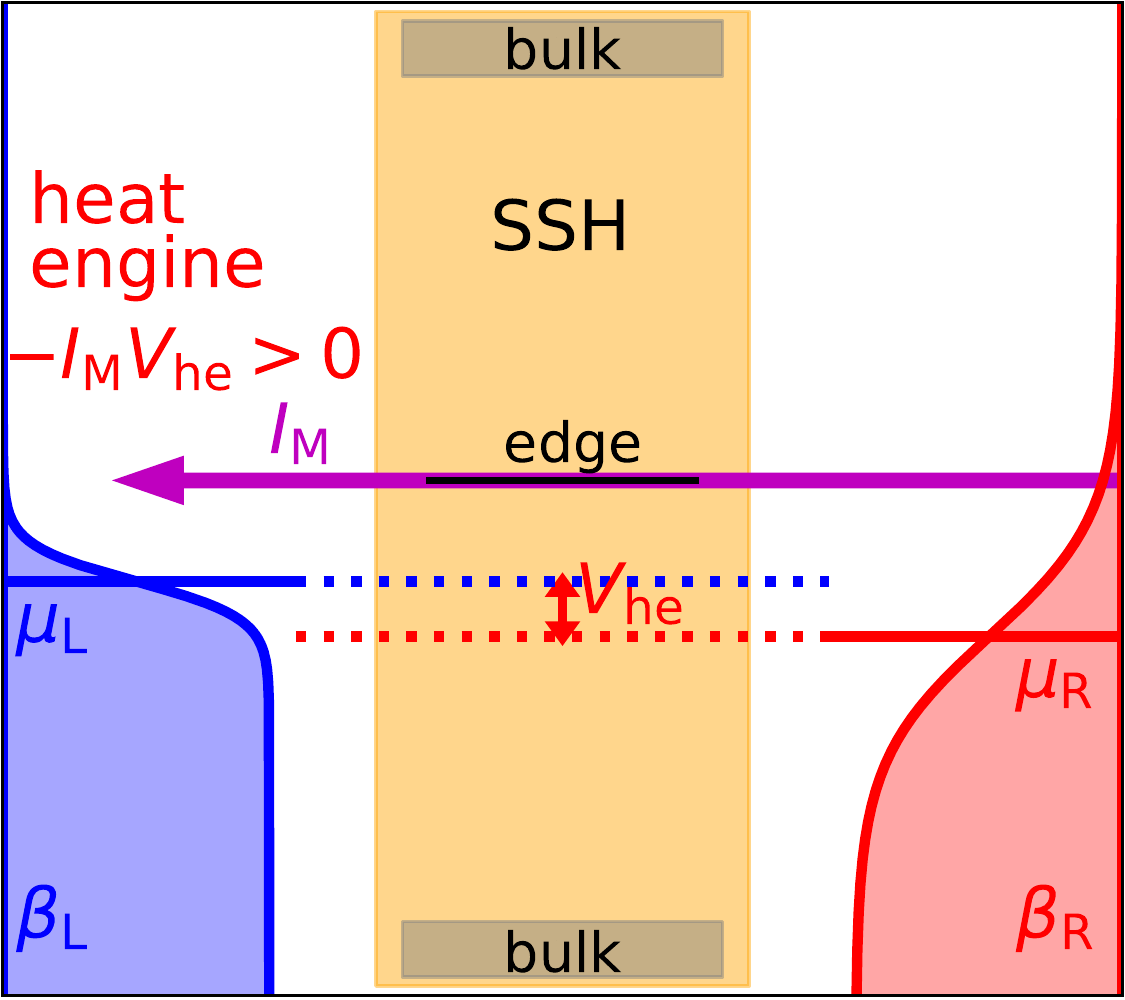}}
	\put(5.,0.){\includegraphics[width=.49\linewidth]{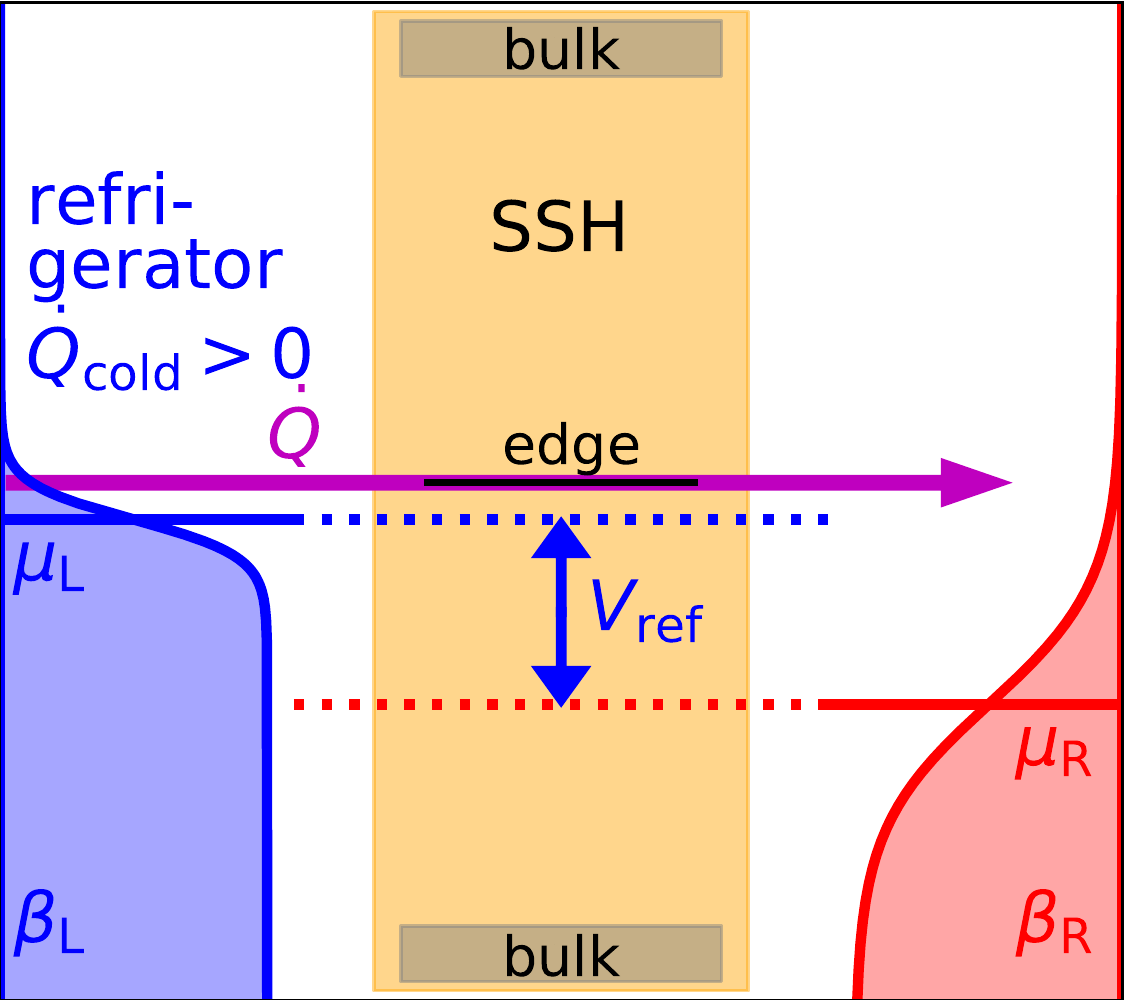}}
	\put(5.1,3.9){\textbf{(c)}}
	\put(0.1,3.9){\textbf{(b)}}
    \put(0.,6.7){\textbf{(a)}}
	\end{picture}
\caption{\label{fig:tightbinding}  
\textbf{(a)} Sketch of a one-dimensional tight-binding chain. 
Alternating nearest-neighbor coupling $\tau_0\pm\delta\tau$ constitutes the $H_{\rm SSH}$ section. 
In addition, the SSH chain is coupled to a left and a right lead $H_{\rm L/R}$ by a reservoir coupling $\tau_{\rm L/R}$.
\textbf{(b)} Thermoelectric transport through the edge states. 
In the heat engine mode, where $-I_{\rm M}V_{\rm he}> 0$, a temperature gradient between left and right lead drives a particle current $I_{\rm M}$ against the 
bias voltage $V_{\rm he}$ creating electric power.
\textbf{(c)} Similar to (b), a refrigerator can be implemented by simply changing the bias voltage to $V_{\rm ref}$ leading to a positive heat current from the left (cold) reservoir into the system $ \dot{Q}_{\rm cold} > 0$
against the thermal gradient.
}
\end{figure}
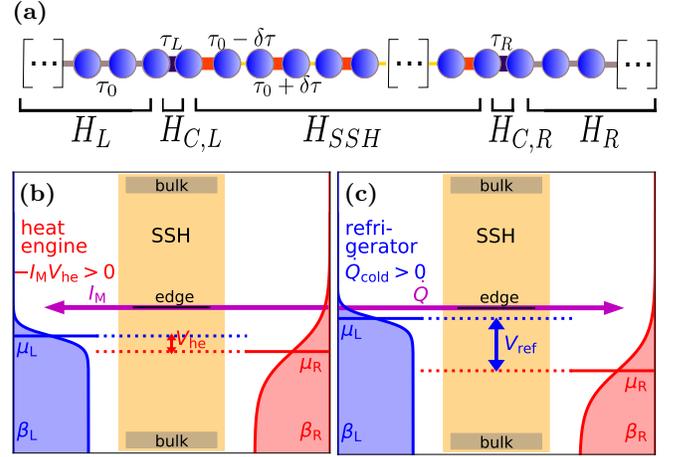

On the other hand, genuine quantum properties have been investigated for their use in heat 
engines~\cite{scully2003b,hardal2015a,manzano2016a}.
In particular, continuous heat engines have the advantage of requiring no external driving or time-dependent coupling
to different reservoirs~\cite{kosloff2014a}.
In contrast, their non-equilibrium environment is generated by multiple equilibrium environments held at
different thermal equilibrium states~\cite{schaller2014}.
The heat entering the system from a work reservoir may e.g. in three-terminal systems be used to cool a cold reservoir~\cite{correa2014a,mu2017a}.
Alternatively, in two-terminal systems chemical work can be extracted from a thermal gradient or can be invested
to cool a cold reservoir~\cite{esposito2009b}.
Thermodynamic efficiencies can then be defined in the usual way by comparing input and output quantities.
Quite generally, the best performance of machines driven by chemical work is found in the so-called {\em tight-coupling limit}, where energy and particle currents 
are proportional~\cite{vandenbroeck2005a,esposito2009c,sanchez2011a}.
In this respect, systems made from few quantum dots have been proposed as energy filters~\cite{jordan2013a}.
However, the required resonance conditions are only met by carefully tuned few level systems, which renders such devices sensitive to perturbations.

The thermodynamic properties of topological materials~\cite{muechler2012a,kempkes2016a} have been
analyzed from the viewpoint of constructing thermoelectric devices~\cite{xu2017a}.
Particularly, this has led to proposals for 2D topological insulators such as 
quantum spin Hall samples~\cite{mani2017a,mani2018a} or 
tailored graphene nanoribbons with heavy adatoms and nanopores~\cite{chang2014a}.
In 1D systems it appears much simpler to maintain thermal gradients between two reservoirs, but 
heat engine perspectives on such systems~\cite{gallego_marcos2017a} have so far not taken topological effects into account.

Within this paper, we therefore propose to employ topological protection against imperfections in a one-dimensional topological insulator 
to realize a robust tight-coupling heat engine and refrigerator.
Specifically, we analyze the nonequilibrium transport and quantum thermoelectric properties of the celebrated Su-Schrieffer-Heeger (SSH) 
model~\cite{Su1979,Heeger1988,Asboth2016} for many non-interacting electrons, but topological properties can 
be found in a larger class of 1D systems~\cite{kitaev2001a,klinovaja2013a,carmele2015a,park2016a,droenner2017a}.
Recently, nonequilibrium transport through an SSH chain has been analyzed in the complementary regimes of strong interactions~\cite{Benito2016,niklas2016a}
and absence of interactions~\cite{Ruocco2017}, and the dynamics of doublons has also been studied~\cite{Bello2016}.
We show that when the SSH chain is attached to two electronic leads, as depicted in Fig.~\ref{fig:tightbinding}, 
its {\em topologically protected edge states
can be selectively addressed to serve as the working medium of a quantum thermal machine}.
Moreover, we show how to exploit a non-equilibrium steady state in order to prepare a situation, where dominantly the topologically induced edge state is occupied, 
while all other states are only weakly populated. 
This finding could be also applied in cold-atom experiments~\cite{Brantut2012,Brantut2013,Jotzu2014,Gross2017,AMMJTDTEI2013,Lohse2015}, where non-trivial topological band structures and related effects have been generated, 
but the preparation of particles in edge modes remains challenging.

The structure of the article is as follows. 
Section~\ref{sec:introNonequ} explains the non-equilibrium setup that will be treated in the following and gives 
an overview on the theoretical methods applied. 
The numerical results are then presented in Sec.~\ref{sec:results}, where transmission, 
site occupation as well as energy and matter currents, noise and Fano factor 
are investigated. 
There, we also discuss the harnessing of boundary modes in a nanothermal heat engine and refrigerator. 
Finally, our conclusions are formulated in Sec.~\ref{sec:conclusion}.


\section{Non-equilibrium SSH model}
   \label{sec:introNonequ}
\subsection{The Hamiltonian}

We consider an effectively one-dimensional system of coupled quantum dots $j$ which are connected to two fermionic leads. 
We describe the system by a tight-binding Hamiltonian as sketched in Fig.~\ref{fig:tightbinding} (a). 
Here, the hopping of spinless fermionic particles 
described by creation operators for each site $\hat c_{j}^\dagger$ is described by the Hamiltonian
\begin{equation}
	H= \sum_{\alpha=\rm R,L}  H_\alpha + H_{\rm SSH} + H_{\rm c},
	\label{eq:hamilton}
\end{equation}
where
\begin{align}
\label{eq:hleft}	H_{\rm L}  &= \sum_{j=-\infty}^{-1} \tau_0 \left( \hat c_{j+1}^\dagger \hat  c_{j} + {\rm h.c.}\right), \\
\label{eq:hright}	H_{\rm R}  &= \sum_{j=N+1}^{\infty} \tau_0 \left( \hat c_{j+1}^\dagger \hat  c_{j} + {\rm h.c.}\right), \\
	H_{\rm c} &=  \tau_{\rm L} \hat c_{1}^\dagger \hat  c_{0} +\tau_{\rm R} \hat  c_{N+1}^\dagger \hat c_{N} + {\rm h.c.}.
\end{align}
Thereby, we have divided the chain into three compartments. 
The central compartment $H_{\rm SSH}$ is denoted as the SSH chain throughout this paper, and it is explained in more detail in App.~\ref{app:isolatedSSH}. 
It is characterized by alternating inter- and intra-dimer couplings, as described by the tight-binding Hamiltonian
\begin{equation}\label{EQ:ham_ssh}
H_{\rm SSH}  = \sum_{j=1}^{N} \varepsilon \hat c_{j}^\dagger \hat  c_{j} + 
\sum_{j=1}^{N-1} \left(\tau_0- (-1)^j\delta\tau \right)\left( \hat c_{j+1}^\dagger \hat  c_{j} + {\rm h.c.}\right)\,,
\end{equation}
with baseline hopping amplitude $\tau_0$ and site-dependent modification $-\tau_0 \le \delta\tau \le +\tau_0$, and an onsite potential $\varepsilon$, 
see the central part of Fig.~\ref{fig:tightbinding}.
Throughout this paper we will only consider the case $N$ even, such that the SSH chain with 
$N$ sites can be interpreted as a chain of $N/2$ dimers.
For odd $N$, the SSH chain has one edge state for both $\delta\tau>0$ and $\delta\tau<0$ (converging to an isolated monomer as $\delta\tau\to\pm \tau_0$).
Therefore,  the character of the phase transition is different in this case, and properties that depend on the edge state overlap 
would not occur for odd sites chains ~\cite{Benito2016}.
Due to the alternating hopping amplitude $t_\pm\equiv\tau_0 \pm \delta\tau$, the  SSH chain exhibits, for long chains, a topological phase transition 
from the {\em topological phase} ($\delta \tau < 0$) towards the {\em trivial phase} ($\delta\tau > 0$).
For short chains, the position of the phase transition still depends on $N$ ~\cite{Delplace2011}, as we discuss in Sec.~\ref{sec:scalability}.

The Hamiltonians $H_{\rm L}$ and $H_{\rm R}$, compare Eqns.~(\ref{eq:hleft}) and~(\ref{eq:hright}), represent the left and right lead, respectively. 
They are coupled to the SSH chain by the coupling Hamiltonian $H_{\rm c}$, where tunnel amplitudes $\tau_{\rm L,R}$ parametrize the coupling strength to the left and right reservoir, respectively.

We stress that in Eq.~(\ref{EQ:ham_ssh}) we have completely neglected particle interactions, which would require to combine our exact Green's function approach 
with perturbative methods. 
This is thus complementary to the treatment in Ref.~\cite{Benito2016}, where the interactions are assumed so strong, such that at most one 
particle can populate the section $H_{\rm SSH}$. 

Our Hamiltonian can in principle be realized in a mesoscopic solid-state setup~\cite{Puddy2015} or molecular wires~\cite{Kocic2015}.
On the other hand, one can also think of a realization of this system in cold-atom experiments. 
The rapid progress in the field of cold atoms allows for the creation of optical potentials with 
almost arbitrary form and time dependence. 
In particular, two-terminal transport setups have been already realized with cold fermions~\cite{Krinner2015,Brantut2012,Brantut2013}. 
In these setups, the interactions can even be tuned via Feshbach resonances, allowing to approach the non-interacting limit.

\subsection{Current and noise}

Throughout the article, we consider a nonequilibrium situation, i.e., we will analyze the transport properties under a finite
DC bias voltage and we will in particular scenarios also consider a thermal gradient.
To this end, we model the total initial density matrix $\rho_{0}$ at time $t=0$ by
\begin{equation}
	\rho_{0} = \prod_{\alpha=L,R} \frac{\exp\left[ -\beta_\alpha (H_\alpha- \mu_\alpha N_\alpha)\right]}{Z_\alpha} \otimes \rho_{\rm SSH}^0\,,
\end{equation}
where $Z_\alpha = \trace{\exp\left[ -\beta_\alpha (H_\alpha- \mu_\alpha N_\alpha)\right]}$ accounts for the normalization.
Here, the operators $N_L = \sum_{j\le 0} c_j^\dagger c_j$ and $N_R = \sum_{j\ge N+1} c_j^\dagger c_j$ count the number of particles in left and right leads, respectively.
Thus, we assume that both leads are initially in a local thermal equilibrium with temperatures $k_{\rm B }T_\alpha = 1/\beta_\alpha$ and chemical potentials $\mu_\alpha$.  
The initial state of the SSH chain is arbitrary, as we are interested in the steady state dynamics for $t\rightarrow \infty$, which does not depend
on $\rho_{\rm SSH}^0$.

The steady-state observables which we investigate in the following are the particle (matter) current through the system $I_{\rm M}$ (counted positive when directed from left to right), 
the corresponding noise $S$ and the occupation $n_j$ of the SSH sites $1 \le j \le N$.
We harness the non-equilibrium Green's function formalism in order to express these quantities in terms of the retarded and advanced Green's functions, which are defined by
\begin{align}\label{eq:gRetAdv}
\mathbf G^{\mathrm r} (E) &= \lim_{\delta\rightarrow  0} \left(E+ \ii \delta \mathbf-  H \right)^{-1}\,,\nn
\mathbf G^{\mathrm a} (E) &= \lim_{\delta\rightarrow  0} \left(E- \ii \delta \mathbf-  H \right)^{-1}\,.
\end{align}
The calculation can be performed in a semi-analytic fashion which we illustrate in Appendix~\ref{APP:greens_function}. 
In terms of the Green's functions~\cite{Economou2006,Haug2008}, the expressions for particle current and noise read
\begin{align}
I_{\rm M}&= \frac{1}{2\pi}\int T(E) \left[f_{\mathrm L}(E) -f_{\mathrm R}(E) \right] dE\,, \label{eq:current}\\
S &= \frac{1}{2\pi}\int \Big\{T(E) \left[ \sum_\alpha f_{\alpha}(E)\left(1- f_{\alpha}(E) \right)\right]\nonumber \\
    &\quad+ T(E) \left[1 -T(E)\right] \left[ f_{\mathrm L}(E) -f_{\mathrm R}(E) \right]^2\Big\} dE\,, \label{eq:noise}
\end{align}
where $f_\alpha(E) = 1/(e^{\beta_\alpha\left(E-\mu_\alpha\right)}+1)$ denotes the Fermi function of lead $\alpha=L,R$. 
These quantities also define the Fano factor $F\equiv S/\abs{I}$ (note that we have absorbed a factor of $1/2$ in the definition of the noise). 
The transmission probability is given by
\begin{equation}
T(E) = \left|G_{1,N}(E) \right|^2 \Gamma_L(E)\Gamma_R(E) \leq 1,
\label{eq:transmission}
\end{equation}
where $G_{i,j}(E)=\left(\mathbf G^{\rm r }(E) \right)_{i,j}$ is the $(i,j)$-th matrix element of the retarded Green's function in position space, and  
$\Gamma_\alpha(E) = \sqrt{4\tau_0^2- E^2} \cdot \tau_\alpha^2/\tau_0^2$ for $-2\tau_0< E<+2\tau_0$ denotes the spectral coupling density of the leads $\alpha$, 
which we explicitly calculate for the chains described by Eqns.~(\ref{eq:hleft}) and~(\ref{eq:hright}) in Appendix~\ref{app:spectralCouplingDensities}. 
This expression for the spectral coupling density $\Gamma_\alpha(E)$ shows that states close to zero energy couple stronger to the reservoirs which will lead in the following to a larger spectral broadening of the edge state compared to the weaker coupled bulk states.

In a similar fashion, one can express the occupation $n_j$ of the SSH chain sites in terms of the lesser Green's function. 
The expression reads
\begin{equation}
n_j\equiv\left<\hat c_j^\dagger(t) \hat c_j(t)\right>_{t\rightarrow \infty} = \int \frac{dE}{2\pi \ii}  G_{j,j}^{<}(E),
  \label{eq:occupation}
 \end{equation}
where the lesser Green's function in position space can be expressed via the retarded and advanced Green's functions as
\begin{align}
	 G_{j,j}^<(E) &= \ii  G_{j,1}^{\rm r}(E) G_{1,j}^{\rm a}(E) \Gamma_L(E) f_L(E)\nn
&\qquad	 +  \ii G_{j,N}^{\rm r}(E)  G_{N,j}^{\rm a}(E) \Gamma_R(E) f_R(E)\,.
\end{align}
Since it follows from Eq.~(\ref{eq:gRetAdv}) that $G_{i,j}^{\rm a} = (G_{j,i}^{\rm r})^*$, we directly see that $n_j\ge 0$.

\subsection{Transport spectroscopy} \label{sec:transpSpec}

\begin{figure}[ht]
 	\centering
 	\setlength{\unitlength}{.1\linewidth}
	\begin{picture}(10,12.2)
	\put(0.,0){\includegraphics[width=\linewidth]{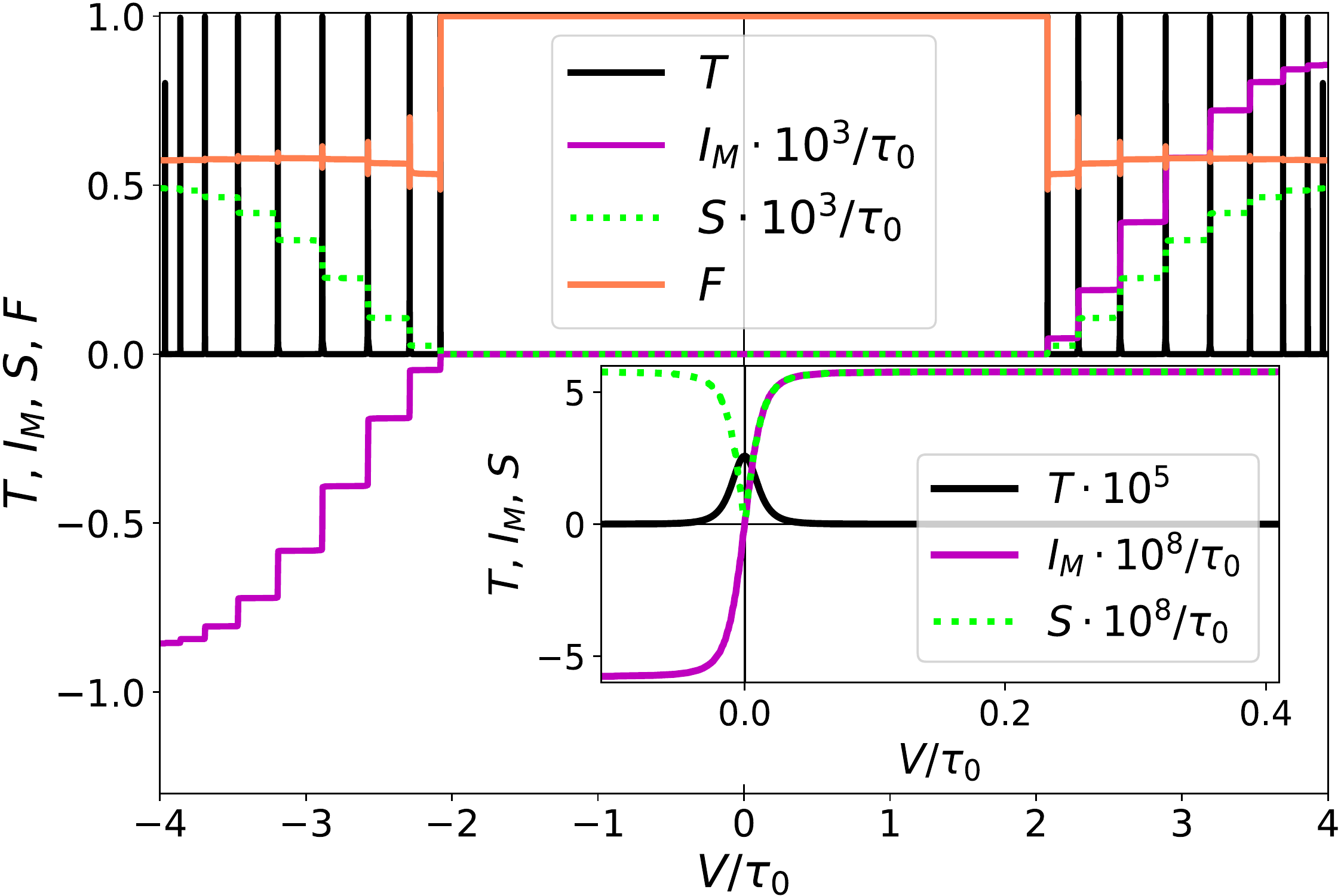}}
	\put(0.,6.9){\includegraphics[width=.45\linewidth]{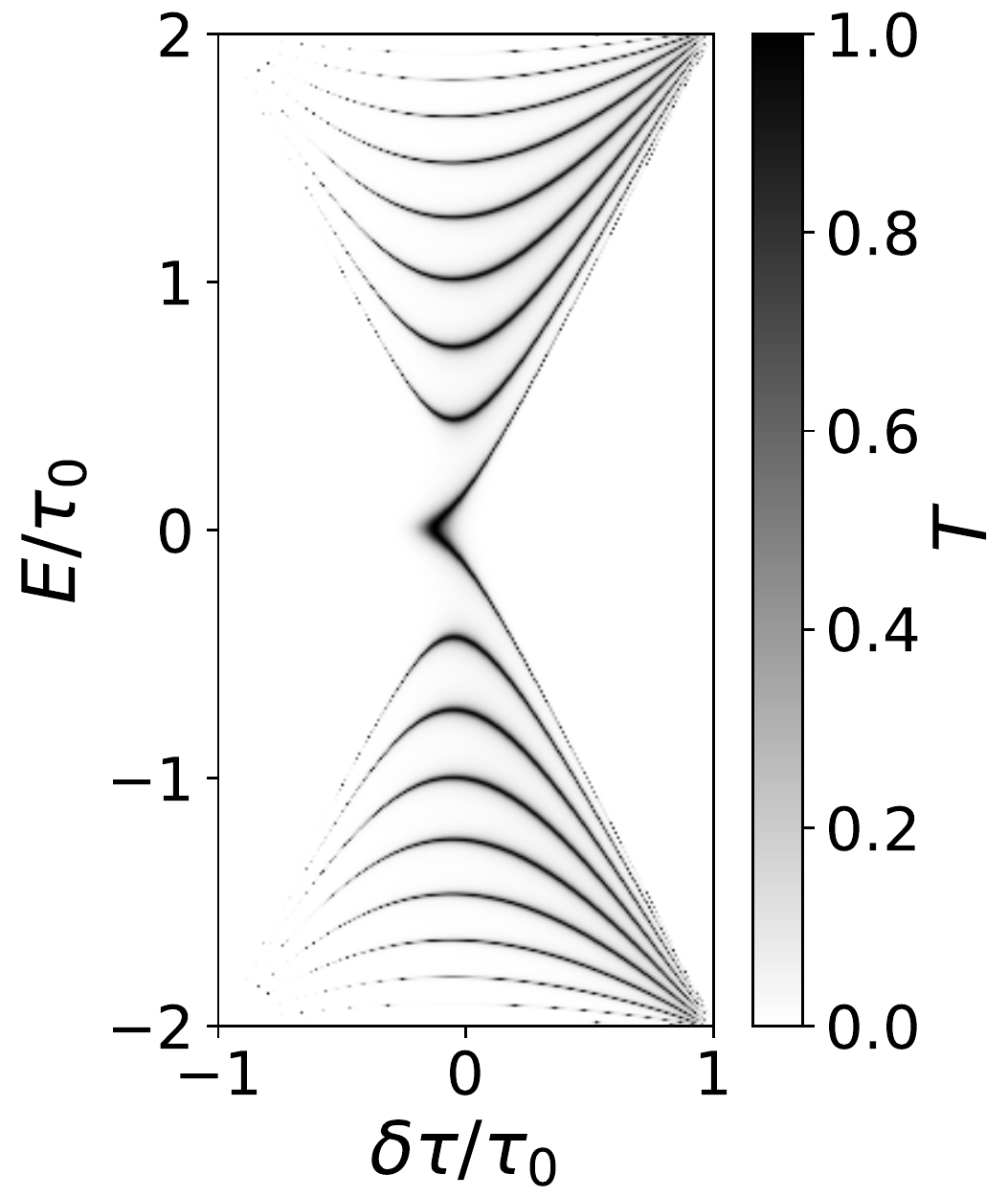}}
	\put(4.8,7.){\includegraphics[width=.5\linewidth]{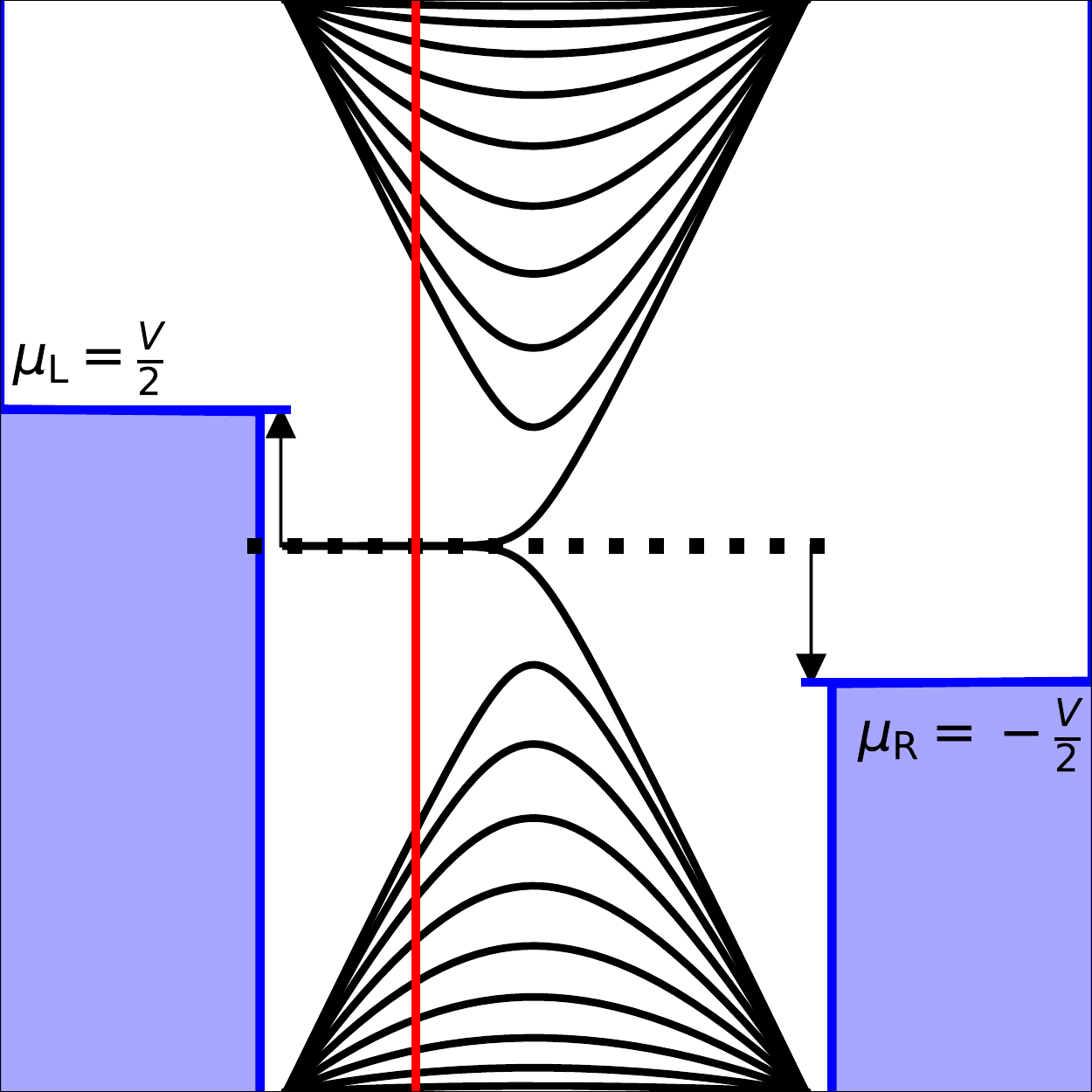}}
	\put(0.,6.8){\textbf{(c)}}
	\put(4.3,12.1){\textbf{(b)}}
    \put(0.,12.1){\textbf{(a)}}
    \end{picture}
    \caption{
    Transport properties for $N=20$ sites ($10$ dimers) at $\varepsilon = 0$. 
    \textbf{(a)} Density plot of the transmission probability $T(E)$ versus energy $E$ and SSH coupling parameter $\delta\tau$ in presence of symmetric reservoir coupling $\tau_L = \tau_R = 0.3\tau_0$.
    \textbf{(b)} 
Sketch of the transport window for 
the left and right reservoir at zero temperature and with chemical potential $\mu_L = +V/2$ and $\mu_R = -V/2$, respectively. 
%
Different chemical potentials in the two leads define the bounds of the transport window, which becomes less sharp at finite temperatures.
The black excitation spectrum -- calculated for $N=20$ sites ($10$ dimers) at $\varepsilon = 0$ -- illustrates that only excitations within the transport window
can participate in transport. 
The red vertical line indicates $\delta\tau/\tau_0 = - 0.5$ used in panel (c).
\textbf{(c)} Transmission $T$, particle current $I_M$, noise $S$, and Fano factor $F$ versus bias voltage $V$ at zero temperature. 
The inset shows a zoom into the central region of the main plot. 
Transmission $T(E)$ is plotted as a function of $V$ by evaluating it at energy $E = V/2=\mu_L=-\mu_R$, at which it 
determines the transport dynamics (compare panel (b)).
\label{fig:spectroscopy}   
}
\end{figure}
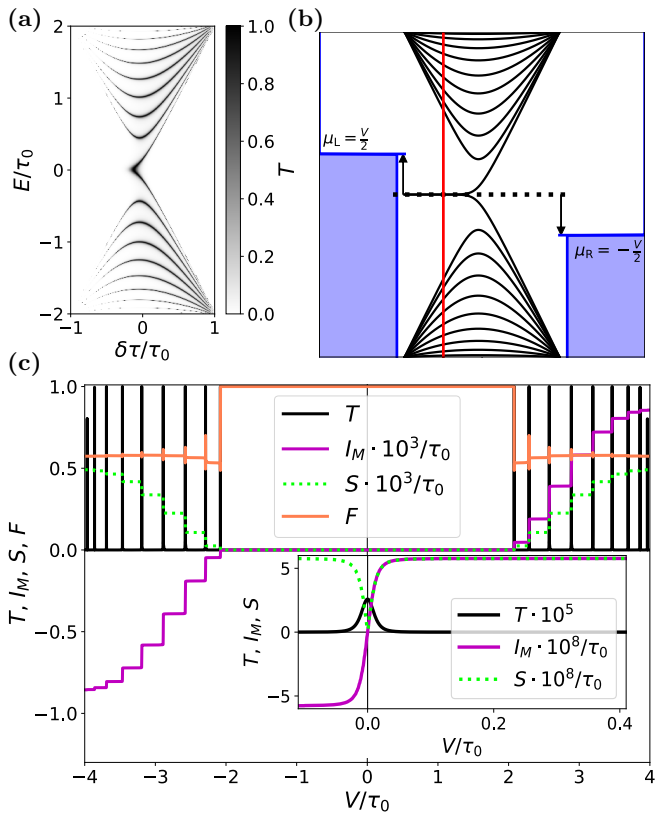

Being the result of an exact calculation, the quantities introduced in the previous section hold for the complete parameter regime.
However, when the couplings $\tau_\alpha$ of the SSH chain to the electronic leads and also the temperatures of the leads are small,
they can even be used for transport spectroscopy, i.e., steps in the currents allow to infer internal parameters of the SSH chain.
In Fig.~\ref{fig:spectroscopy} (a) the transmission is plotted as a function of energy and topological control parameter $\delta \tau$.
We observe that the maxima  of $T$  are closely related to the energy spectrum of the isolated chain shown in Fig.~\ref{fig:spectroscopy} (b).
One also observes that the transmission peaks located in the topologically trivial phase at $\delta \tau >0$ are more intense
than their topologically nontrivial counterparts at $\delta\tau<0$, which can be expected to cause higher particle and energy currents in the first case. 
Furthermore, the transmission through the exponentially localized midgap states vanishes quickly for $\delta \tau \ll0$, leading to one pair of transmission peaks less compared to the trivial phase. 

By introducing a finite bias $V = \mu_L-\mu_R \neq 0$, we can open a {\em transport window} consisting of energy values at which states are occupied in one lead and empty in the other. 
This is shown in Fig.~\ref{fig:spectroscopy} (b). 
When we now vary the bias voltage, the current through the SSH chain changes 
as additional eigenstates of the SSH chain enter the transport window. 
Figure~\ref{fig:spectroscopy}(c) depicts the particle current $I_M$, noise $S$, and Fano factor $F$ as a function of bias $V$, as well as the transmission $T(E)$ as a function of $E = \mu_L=V/2$. 
These calculations were performed in the topological phase corresponding to a cut through Fig.~\ref{fig:spectroscopy} (b) along the red line, and at zero temperature where the Fermi functions transform into Heaviside-$\Theta$ functions. 
The inset shows a tiny transmission peak for the edge state at $E=0$ which leads to a small but finite current as soon as we open the transport window around $V=0$. 
From the sole height of the peak one would expect the associated edge state current to be vanishingly small compared to 
the current supported by the bulk states.
However, at large enough bias, the current is given rather by the integral over the transmission, such that the total area under the
transmission peak becomes relevant.
Therefore, for a broadened edge state transmission, the associated current is not as drastically suppressed
as one might naively expect from the transmission height.
The sign of the current confirms that the particle transport is directed from the lead with the higher chemical potential to the one with the lower chemical potential. 
Current as well as noise remain constant until $V$ reaches a value where another transmission peak enters the transport window leading to a step in the current and noise.  
Although these calculations have been performed at zero temperature, 
we observe a small but finite broadening of the current 
and noise steps, resulting from the finite width of the transmission peaks due to the 
finite coupling strengths with the contacts.
A simple master equation approach would not be able to include these effects.

\subsection{Thermodynamic properties}\label{sec:introEq}
 
In absence of explicit driving, one has to apply both, temperature and potential gradient to use the device as a thermoelectric generator or refrigerator.
Without loss of generality, we will consider a cold left lead and a hot right lead 
$T_{\rm L} < T_{\rm R}$ ($\beta_{\rm L}>\beta_{\rm R}$) in addition to a voltage bias 
$V= \mu_{\rm L} - \mu_{\rm R}$.
We consider the electric power
\begin{equation}
P = - I_M \cdot V\,,
\end{equation}
where a positive power $P>0$ corresponds to a matter current against the bias, which can be used for different applications (e.g., charging a battery), 
while a negative power $P<0$ is dissipated as heat to the reservoirs. 
Power generation ($P>0$) is driven by the heat current entering the SSH chain from the hot (right) 
reservoir $\dot{Q}_{\rm hot}$, which for our conventions becomes
\begin{equation}\label{EQ:heatcur_hot}
  \dot{Q}_{\rm hot} = -(I_{\rm E} - \mu_R I_{\rm M})\,.
\end{equation} 
Here, the stationary energy current $I_{\rm E}$ traversing the SSH chain from left to right reads
in analogy to Eq.~(\ref{eq:current})
\begin{equation}\label{EQ:current_energy}
I_E = \frac{1}{2\pi} \int T(E) \cdot E \left[ f_L(E) - f_R(E)\right] dE\,,
\end{equation}
and the sign in front of Eq.~(\ref{EQ:heatcur_hot}) results from the convention that currents count positive when they
enter the system, whereas $I_E$ and $I_M$ are positive when directed from left to right.
In the same way we define the heat current from the cold reservoir
\begin{equation}
\dot{Q}_{\rm cold} =  I_E - \mu_L I_M.
\end{equation}
These definitions ensure that positive heat currents \mbox{$\dot{Q}_{\alpha} > 0$} describe processes where heat from reservoir $\alpha$ enters the dimer chain, while negative heat currents indicate that heat is leaving the SSH chain towards reservoir $\alpha$. 
If the particle current vanishes, the heat currents will equal the energy current up to a sign.

\textbf{Heat engine.} 
For certain parameters, the dimer chain generates power $P>0$ due to heat entering the system from the hot reservoir. 
In this case, one observes a particle current against the bias voltage. 
For non-interacting electronic transport, a Landauer representation of energy (see Eq.~(\ref{EQ:current_energy})) and matter (see Eq.~(\ref{eq:current})) 
currents is generic, and in this case one can demonstrate the second law of thermodynamics at steady state~\cite{nenciu2007a,Topp2015,yamamoto2015a}.
The second law forbids that the heat flow from the hot reservoir $\dot{Q}_{\rm hot}$ is completely 
transformed into electric power. 
In particular, the efficiency defined by the ratio of generated electric power and heat flow from the hot reservoir 
\begin{equation}
\eta = \frac{P}{\dot{Q}_{\rm hot}} \Theta(P) 
= \frac{I_{\rm M} V \Theta(-I_{\rm M} V)}{I_{\rm E}-\mu_R I_{\rm M}} \leq \eta_{\rm C}\,,
\end{equation}
is bounded by Carnot efficiency 
\bea
\eta_{\rm C} = 1-\frac{T_{\rm L}}{T_{\rm R}} = 1 - \frac{T_{\rm cold}}{T_{\rm hot}} < 1\,.
\eea
Under normal circumstances, Carnot efficiency can only be achieved under somewhat pathological circumstances:
To reach the equality, the heat engine has to operate without entropy production, which practically means
that e.g. for cyclic heat engines, it can only be reached by infinitely slow (adiabatic) evolutions, see e.g.~\cite{gelbwaser_klimovsky2013a}.
Similarly, for continuously operating heat engines~\cite{kosloff2014a}, this means that the power output must vanish.
Therefore, it is more customary to consider the efficiency at finite power output, which usually is a sophisticated numerical optimization problem.

\textbf{Refrigerator.} 
Another thermodynamic application can be obtained by tuning the parameters such that the heat current $\dot{Q}_{\rm cold}$ becomes positive. 
In this regime, heat is entering the dimer chain from the cold reservoir, effectively cooling it. 
This process can only be achieved by $P<0$, i.e., investing work (with a particle current along the bias voltage). 
The ratio of heat leaving the cold reservoir and invested chemical work defines the coefficient of performance (COP) of the cooling process
\begin{equation}
{\rm COP} = - \frac{\dot{Q}_{\rm cold}}{P} \Theta(\dot{Q}_{\rm cold}) \le {\rm COP}_{\rm C} \,.
\end{equation}
The second law bounds it by the Carnot value
\begin{equation}
{\rm COP}_{\rm C} = \frac{\beta_R}{\beta_L - \beta_R} = \frac{T_{\rm cold}}{T_{\rm hot}-T_{\rm cold}}\,.
\end{equation}

Since the COP can exceed one, we will renormalize efficiency $\eta$ and COP by their maximum Carnot values.


 \section{Results} 
  \label{sec:results}

\subsection{Site occupation}

\label{sec:edgeStatePreparation}

The objective of this section is to show how to exclusively prepare a situation, where only  the edge states are  occupied, while the bulk states are empty. 
In measuring then the local occupations of all SSH sites, one could prove the existence of the topologically protected modes, which in our case turns out to be a 
superposition of midgap states that is localized to the left end of the chain.

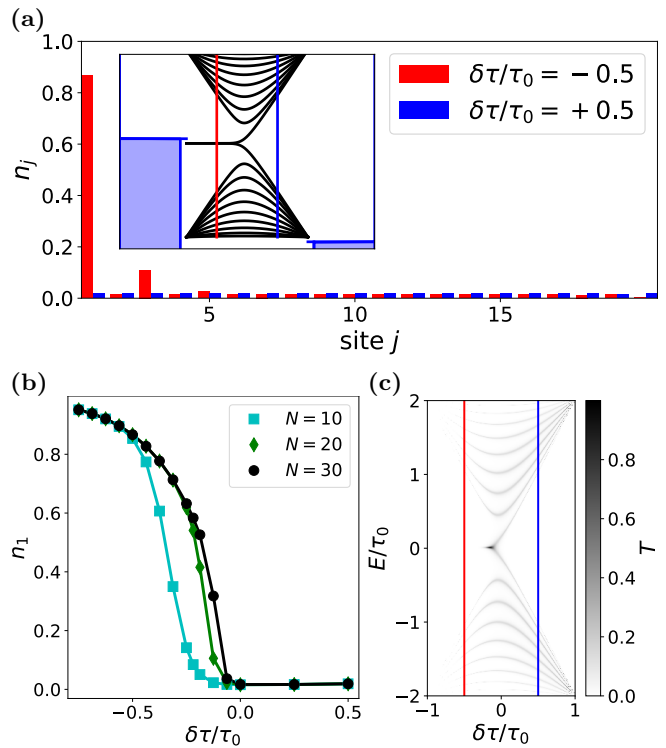
\begin{figure}[ht]
	\centering
 	\setlength{\unitlength}{.1\linewidth}
	\begin{picture}(10,11.2)
	\put(0.,0){\includegraphics[width=.54\linewidth]{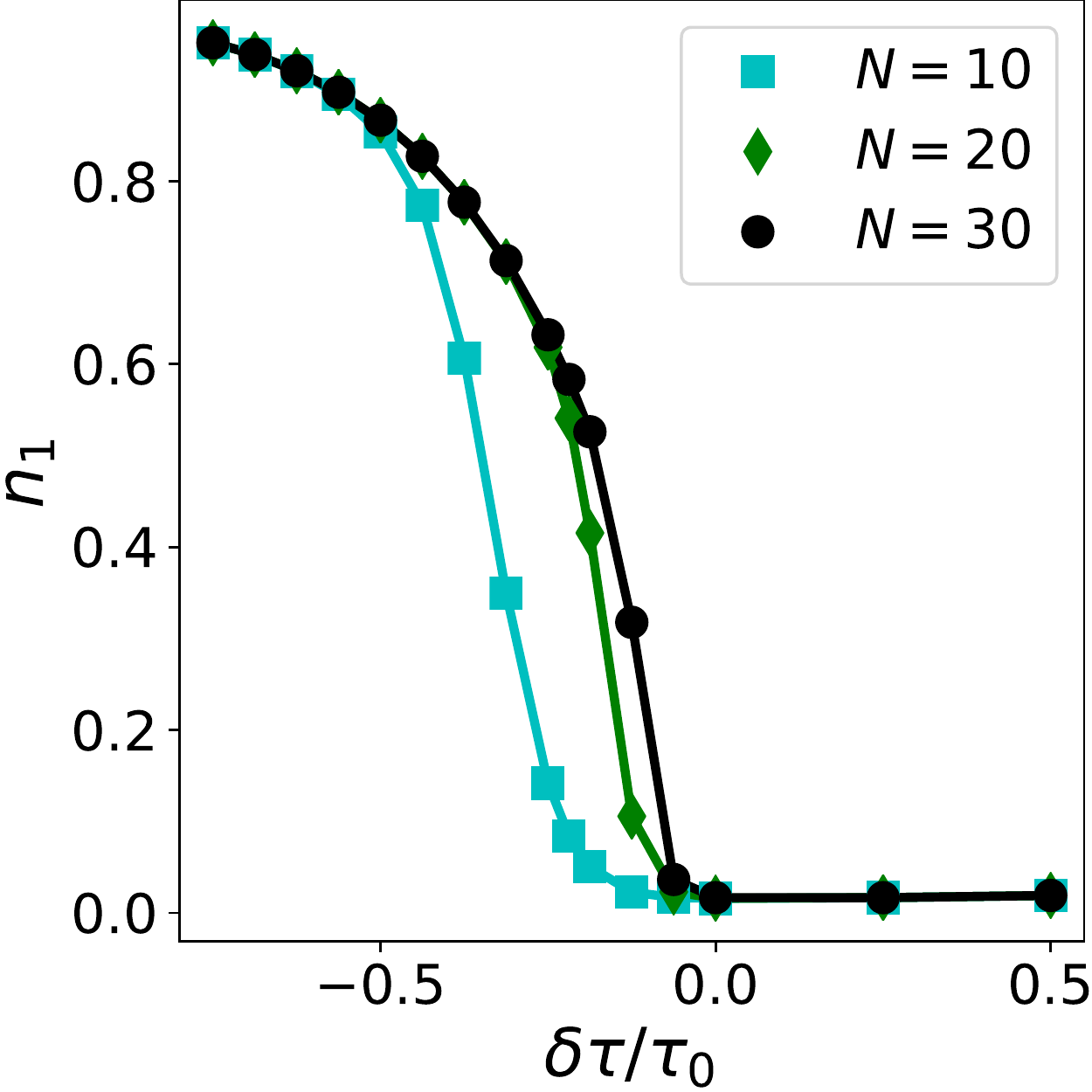}}
	\put(0.,6.){\includegraphics[width=\linewidth]{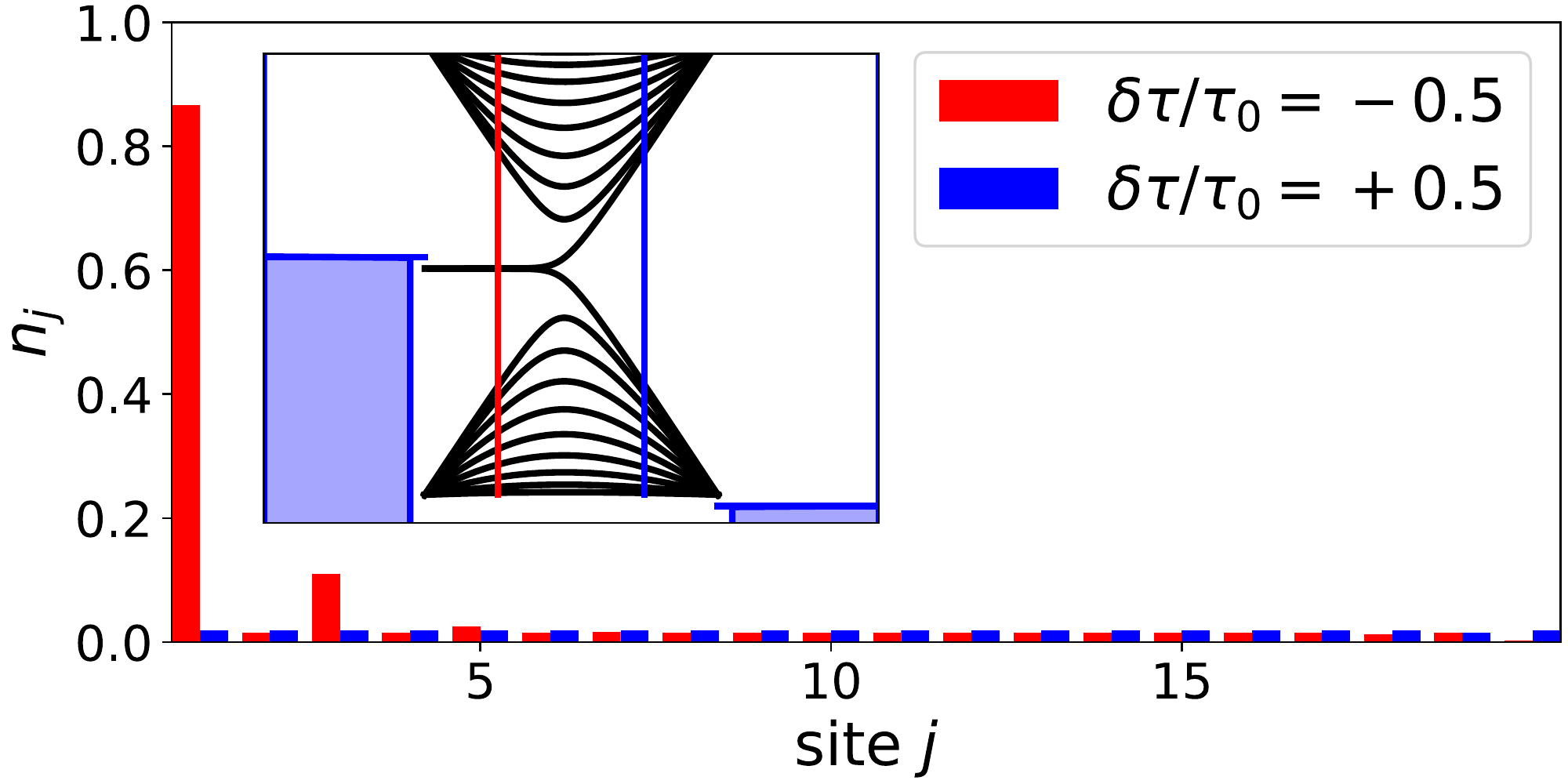}}
	\put(5.5, 0.){\includegraphics[width=.44\linewidth]{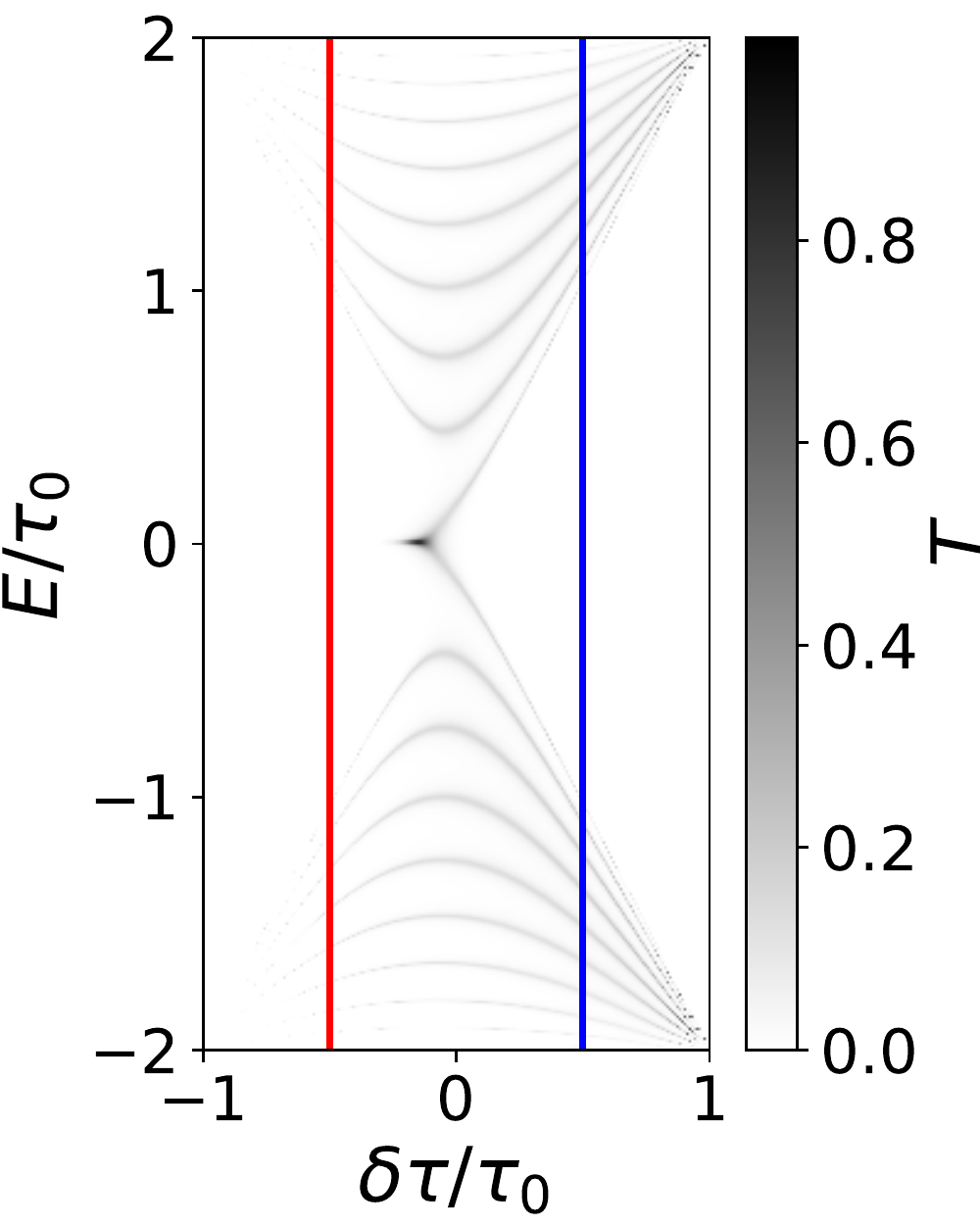}}
	\put(5.5,5.5){\textbf{(c)}}
    \put(0.,5.5){\textbf{(b)}}
    \put(0.,11.1){\textbf{(a)}}
	\end{picture}
	\caption{Stationary site occupations and transmission for asymmetric reservoir couplings $\tau_L/\tau_0 = 0.1$ and $\tau_R/\tau_0 = 0.5$, chemical potentials $\mu_L/\tau_0 = 0.1$, 
	$\mu_R/\tau_0 = -2.1$, and on-site energy $\varepsilon=0$ at zero temperature.
	\textbf{(a)} Stationary occupations of sites $j$ for a SSH chain consisting of $10$ dimers in the topological (red) an trivial (blue) phase. 
	\textbf{(b)} Stationary occupation $n_{j=1}$ of the first site of a SSH chain of different lengths as a function of  the SSH coupling 
	parameter $\delta\tau$. 
	The transition becomes sharper as $N\to\infty$, demonstrating the edge state manifestation during the normal-to-topological 
	phase transition.
	\textbf{(c)} Transmission $T$ as a function of energy $E$ and SSH coupling parameter $\delta\tau$ for $\varepsilon=0$ and the given asymmetric reservoir coupling. 
	Red and blue lines indicate the values of $\delta\tau$ used in (a).
	}
    \label{fig:occupation}   
	\end{figure}

To achieve this, we weakly couple the SSH chain to the left lead (acting as {\em source}), while we apply a stronger coupling to the right lead 
(acting as {\em drain}) $\tau_L < \tau_R < \tau_0$. 
The last constraint is needed to ensure for an approximate validity of an intuitive master equation model. 
Moreover, the chemical potential of the weakly coupled left reservoir is adjusted to be energetically above the midgap modes $\mu_L > \varepsilon$,
while the chemical potential of the strongly coupled right lead shall be below the lowest valence band state $\mu_R < - 2 \tau_0$.
This is sketched in the inset of Fig.~\ref{fig:occupation} (a).
Here, the asymmetry of the reservoir couplings, in addition to the described bias voltage configuration, is needed for the reduction of the bulk state contribution to the current.

The resulting occupations  $n_j$ of the different sites $j$ according to Eq.~\eqref{eq:occupation} are depicted in Fig.~\ref{fig:occupation} (a). 
Importantly, we observe that the occupation in the topological phase ($\delta \tau/\tau_0 =-0.5$) strongly resembles (half of) the wave function of the isolated SSH chain 
depicted in red in Fig.~\ref{fig:spectrum} (see Appendix~\ref{app:isolatedSSH}). 
In particular, the vanishing occupation of the even sites and the exponential decay towards the center of the SSH chain signify that the present choice of parameters can 
be exploited to prepare a superposition of midgap states $\left|L\right> =\frac{1}{\sqrt{2}} \left( \left|+\right> + \left|-\right> \right)$, which is localized to the left side of the chain, 
with high fidelity.
Moreover, Fig.~\ref{fig:occupation} (a) shows that the occupation of the bulk states is negligible in the topological phase as desired.
The absence of bulk occupation is not trivial, because even in a small transport window that excludes the bulk state contribution to the current, we would in general still observe an occupation of the valence band.
This occupation of the bulk normally results from the fact that for a small positive bias voltage, electrons enter from the left reservoir into the valence band, but get trapped there due to a lack of free states at corresponding energies in the right reservoir.
 
The situation in Fig.~\ref{fig:occupation} (a) 
can be explained as follows:  An electron entering from the source on the left into the left edge state $\left|L\right>$ becomes 
trapped inside the SSH chain due to the state's localization far away from the drain. 
It cannot return to the source as the source reservoir modes are almost completely occupied.
If, by chance, the right edge state $\left|R\right>$ would be occupied (from left or right lead), it would be quickly emptied dominantly 
to the drain as its states are all empty and strongly coupled. 
In contrast, due to the assumed coupling asymmetry, electrons entering a bulk state from the left reservoir quickly propagate to the right end of the chain where  
they exit into strongly coupled drain. 
In this way, we can avoid the additional occupation of bulk states, paying the price that also the occupation of the edge state localized to the right end $\left|R\right>$ 
vanishes as an electron in this state $\left|R\right>$ would also leave the chain quickly into the strongly coupled reservoir.

In addition, we compute the occupation of the first site of the SSH chain that hosts the maximal occupation in the topologically nontrivial phase. 
Fig.~\ref{fig:occupation} (b) depicts this quantity for different SSH chain lengths versus coupling parameter $\delta\tau$. 
We observe  that the occupation $n_1$ of site $j=1$ is finite in the topologically non-trivial phase $\delta\tau<0$, 
but rapidly disappears when approaching and crossing the topological phase transition at $\delta\tau=0$.  
This observable thus constitutes a clear order parameter signaling the topological phase transition of the system in this particular nonequilibrium setup. 

Moreover, the site occupation $n_1$ in the non-trivial phase also depends on the chain length $N$. 
As Fig.~\ref{fig:occupation} (b) suggests, the crossover at $\delta \tau =0$ becomes sharper for longer chain lengths. 
Indeed, the phase transition in chains of finite dimer numbers $M$ rather takes place at~\cite{Delplace2011} $t_+/t_-= 1-1/(M+1)$. 
In the infinite chain limit, this would give rise to a non-analyticity of $n_1$ at the topological phase transition. 

The transmission for the mentioned asymmetric reservoir couplings which we exploited for the described edge state preparation is shown in Fig.~\ref{fig:occupation} (c).
In contrast to the symmetric case, we observe that the transmission for a stronger coupling at one lead becomes broader (not explicitly shown) and has a lower intensity as compared to the 
symmetric case depicted in Fig.~\ref{fig:spectroscopy} (a).
An exception is the region around $E=0$, where two bulk states are transformed into exponentially localized midgap states 
(we show the wave function of one of them in the appendix in Fig.~\ref{fig:spectrum}). 
Thus, asymmetric reservoir couplings seem to have a tremendous impact on the bulk states. 

Finally, we  note that the occupation is experimentally accessible by harnessing an adjacent quantum point contact.
After reaching the stationary state, one can measure the occupation of the site $j=1$, which can take values 
$\tilde n_1=\left\lbrace 0,1 \right\rbrace$. 
Averaging over plenty of these experimental runs, one can determine the mean occupation $n_1$. 

\subsection{Current and noise at finite temperatures}
    
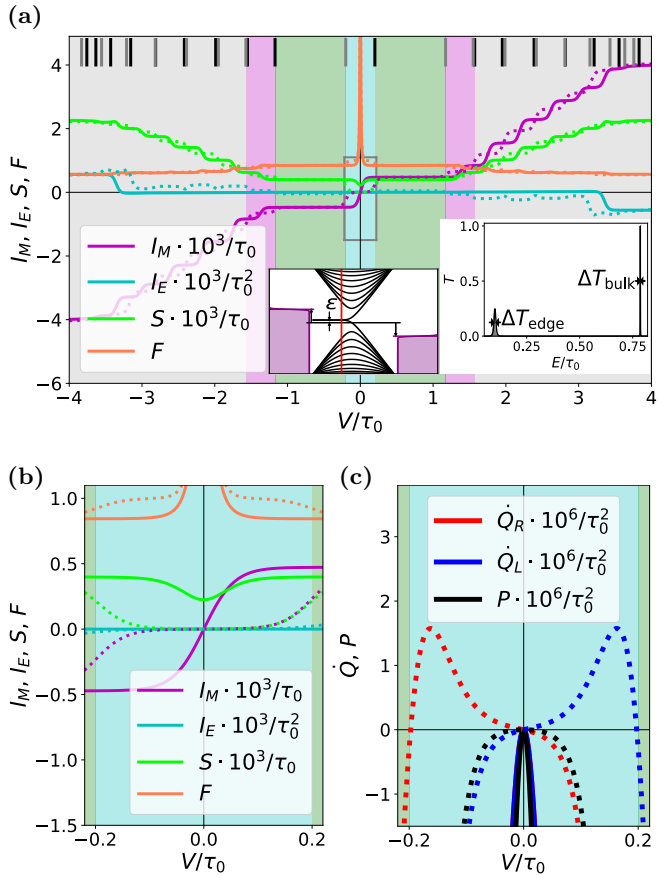
\begin{figure}[ht]
    \centering
     \setlength{\unitlength}{.1\linewidth}
	\begin{picture}(10,13.1)
	\put(0.,6.8){\includegraphics[width=1.\linewidth]{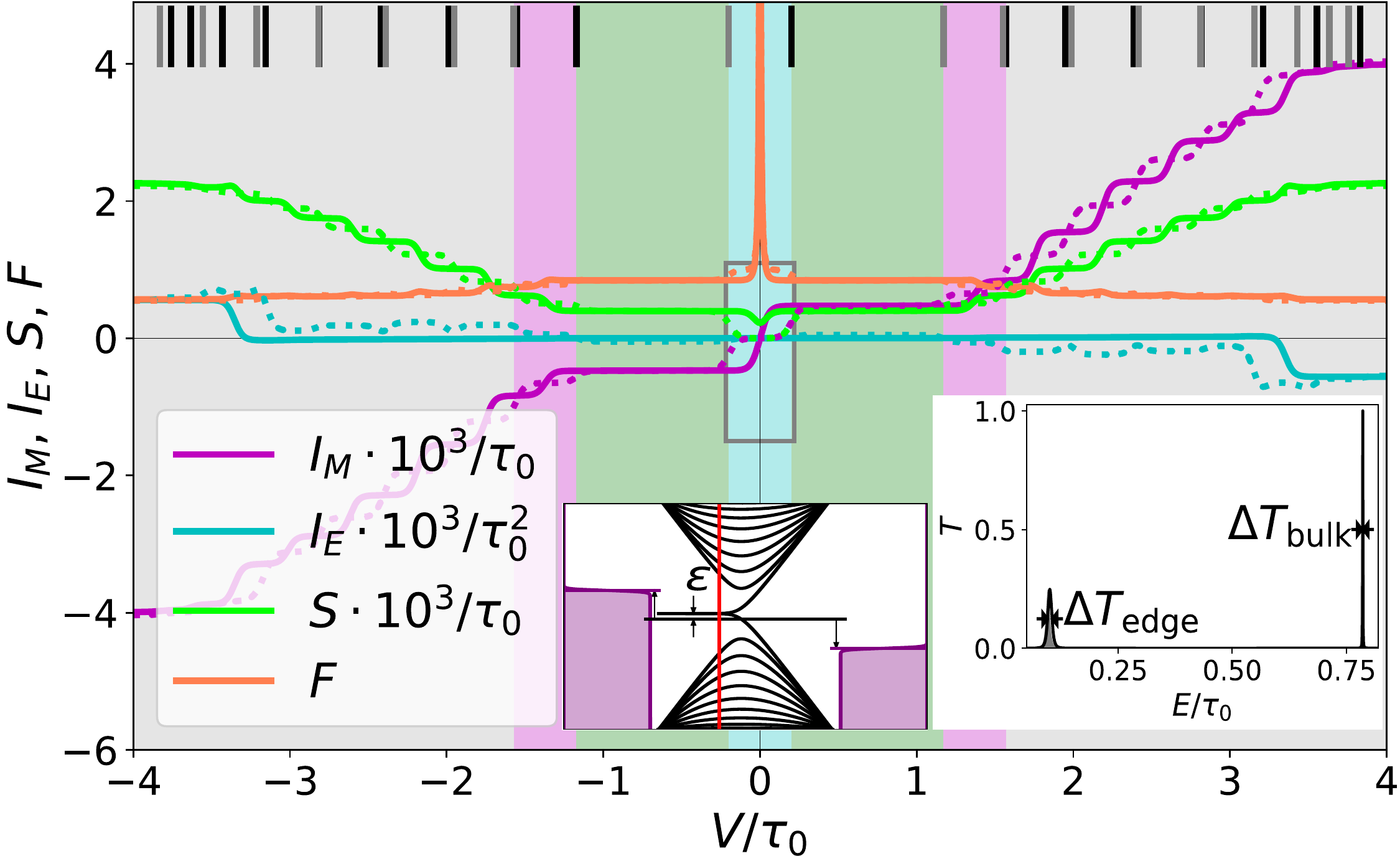}}
	\put(0.,0.){\includegraphics[width=.49\linewidth]{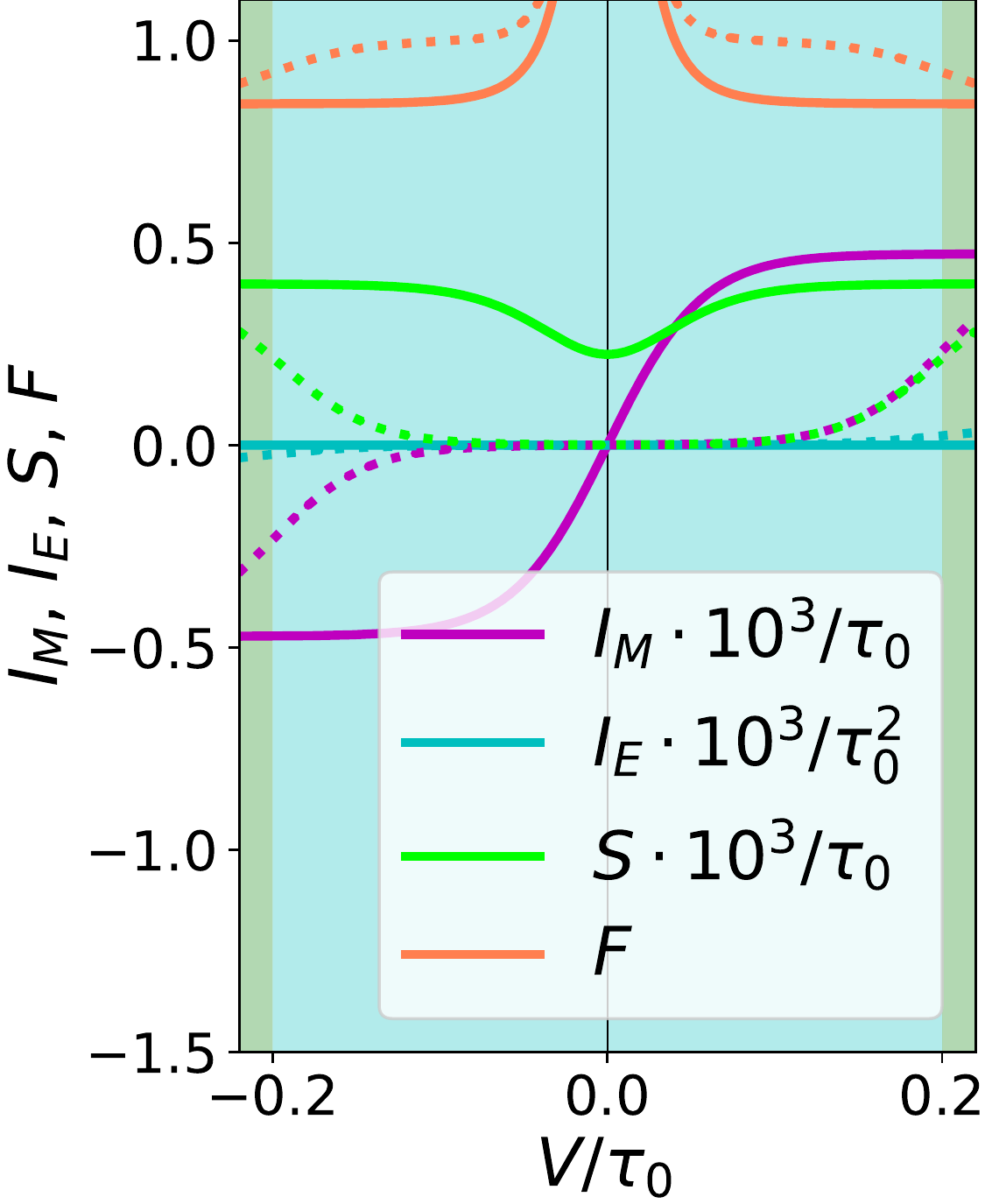}}
	\put(5.,0.){\includegraphics[width=.49\linewidth]{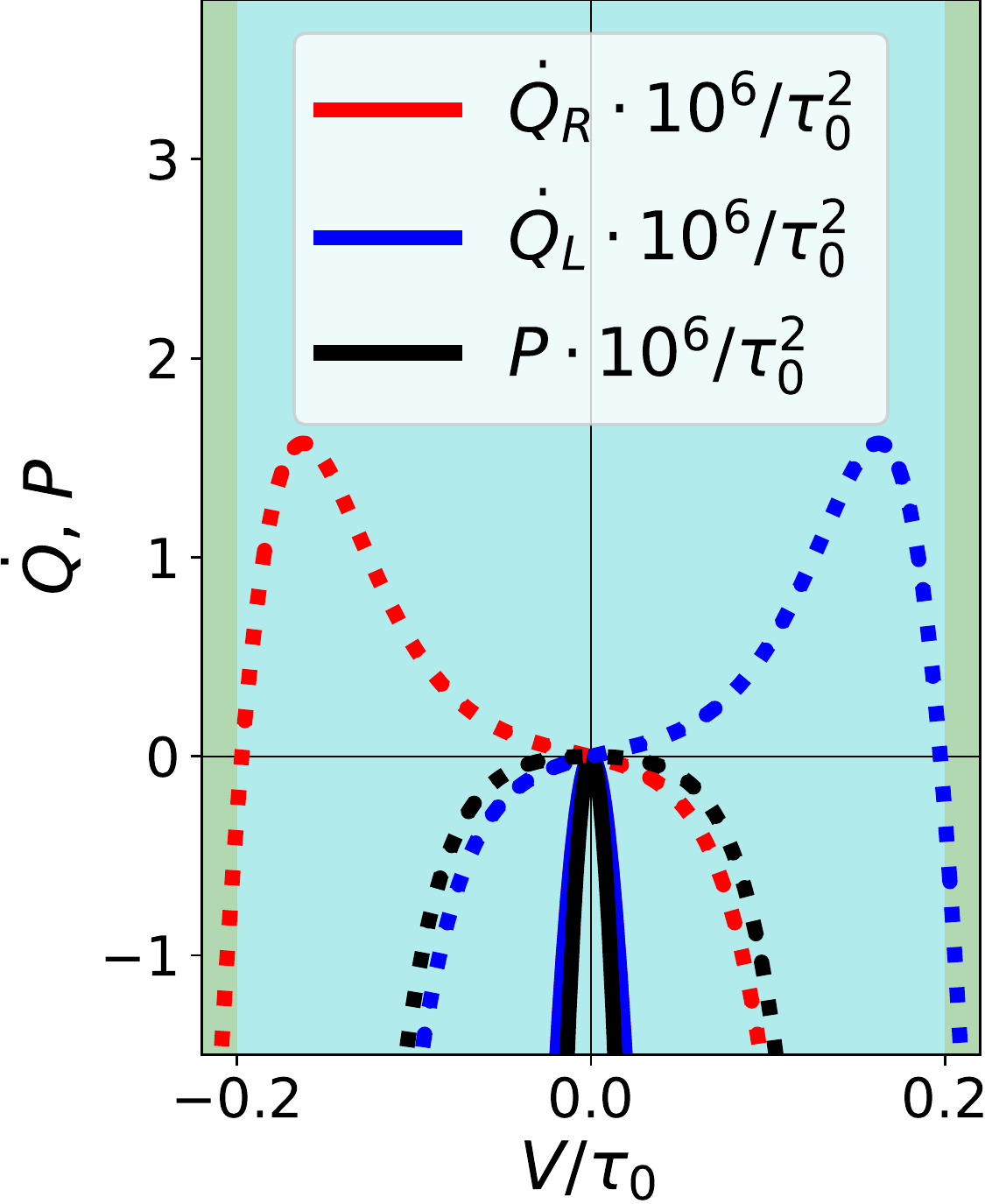}}
	\put(0.,13.1){\textbf{(a)}}
	\put(0.,6.){\textbf{(b)}}
	\put(5.1,6.){\textbf{(c)}}
	\end{picture}
   \caption{Transport at equal reservoir temperatures $\beta_L = \beta_R = 75/\tau_0$. 
   Solid lines correspond to results without any onsite potential, while dotted lines refer to a system where onsite energy $\varepsilon = 0.1\tau_0$. Calculations were performed for $10$ dimers in the topological phase $\delta\tau/\tau_0 = -0.3$, and at reservoir couplings $\tau_L = \tau_R = 0.1\tau_0$.
 \textbf{(a)} Particle current $I_M$, energy current $I_E$, noise $S$ and Fano factor $F$. 
 Black (grey) lines in the upper part of the figure indicate the position of the transmission peaks as a function of $E=\mu_L=+\frac{V}{2}$ ($E=\mu_R=-\frac{V}{2}$), only for the $\varepsilon=0.1\tau_0$ case. The inset in the middle illustrates the transport window opened by the two reservoirs at equal, finite temperatures, for $\varepsilon = 0.1\tau_0$ which corresponds to the dashed lines. 
 The right inset shows the transmission of edge and first bulk states for the parameters of the dotted lines in the main plot and
 demonstrates that the width of the edge state transmission is significantly larger.
   \textbf{(b)} Zoom into the part of (a) where the dynamics is dominated by the midgap states.
 \textbf{(c)} Heat currents $\dot{Q}$ from left and right reservoir, and power $P$. 
 In the absence of onsite energy $\varepsilon = 0$ (solid curves), the heat currents of both reservoirs are identical.
 }
    \label{fig:equalTemperature}   
    \end{figure}

In this section, we will consider equal, finite temperatures on both leads and particularly investigate the behavior of current and
noise in the topological phase.
Since we will need to consider finite edge state energies $\varepsilon$ later, Fig.~\ref{fig:equalTemperature} compares the transport properties of a system with $\varepsilon=0$ (solid lines) to a situation where a potential $\varepsilon=0.1\tau_0$ is applied to the sites of the SSH chain (dotted lines).
In contrast to the previous section, we will limit the following calculations to symmetric reservoir couplings. However, one can show that asymmetric reservoir coupling leads to qualitatively very similar results.

First of all, we see in Fig.~\ref{fig:equalTemperature} (a) that at sufficiently low temperatures, 
current and noise show the usual steps as a function of bias voltage, similar to Sec.~\ref{sec:transpSpec}.
However, the steps now appear smoother due to finite temperatures in the leads. 
This shows that transport spectroscopy can be used to 
experimentally determine the excitation spectrum of the central SSH chain. 
In particular, we can detect the topological phase by the small but finite matter current mediated by the edge
states at any small bias voltage (solid curves), which can be observed in Fig.~\ref{fig:equalTemperature} (b).
Here, we see that the edge state current is comparable to the current mediated by the bulk states although the height of the edge state transmission 
is significantly smaller. 
However, we have also checked the integral over the corresponding peaks of the transmission, which are in good agreement with the
currents supported by edge and bulk states.
This is illustrated in the right inset of Fig.~\ref{fig:equalTemperature} (a) where we observe a significantly broader transmission peak for the edge state in comparison to the first bulk state. Due to our choice $\delta \tau /\tau_0 = 0.3$, we get closer to the topological phase transition than in Fig.~\ref{fig:spectroscopy} (c), which explains the higher edge state transmission.
When we choose finite $\varepsilon$ (dotted curves), we see that a region of vanishing matter current opens up (turquoise background), which
is expected as by tuning $\varepsilon$ we simply shift all the excitation energies of the SSH chain and only leave the
vacuum energy (empty chain) invariant.
The central inset of Fig.~\ref{fig:equalTemperature} (a) sketches this shift.
Further increasing $V$, the edge states start to participate in the transport (green region), until the bulk states enter the transport window (purple).
In the large bias regime (grey), further bulk states enter.
For infinitely long SSH chains (sufficiently large $N$), one could no longer resolve any current steps in this 
regime as the bands become continuous.

The zoom in Fig.~\ref{fig:equalTemperature} (b) shows that although the midgap modes participate in transport (purple line), 
for $\varepsilon=0$ (solid)  they do not transport any energy.
This is different for finite $\varepsilon$ (not explicitly shown in the figure, but can be seen from Eq.~(\ref{EQ:current_energy})) where we observe a small but finite energy current inside the band gap. As the energy current results from the integral shown in Eq.~(\ref{EQ:current_energy}), it has to yield small values inside the band gap because the already relatively small edge state transmission vanishes quickly as energy increases. However, the obtained $I_{\rm E}$ is sufficient to enable later applications e.g. in a heat engine, compare Sec.~\ref{sec:thermoelectric}.

We observe that the noise (solid green) is positive throughout, as it has to be.
At vanishing bias voltage, it solely results from the thermal fluctuations of the leads, and correspondingly the Fano factor (solid orange) 
diverges there.
For $\varepsilon=0$, the Fano factor immediately drops slightly below one as soon as the bias voltage is increased, 
indicating sub-Poissonian transport and a slight anti-bunching of electronic counting statistics.
In contrast, we see in Fig.~\ref{fig:equalTemperature} (b) for finite $\varepsilon$ (dotted) that the Fano factor first roughly drops to a small plateau where $F\approx 1$, before further relaxing to the $\varepsilon=0$ value as the edge states begin to participate in transport, thereby uniting with the solid curves at the boundaries.
When transport is non-negligible, the sub-Poissonian noise level can be understood as due to Pauli-blocking, anti-bunching between subsequently tunneling electrons builds up.
In contrast, the value $F\approx 1$ can be understood as for finite $\varepsilon$ the edge states in this bias window are hardly ever occupied 
(as the edge state transmission peak is energetically above the chemical potentials of both reservoirs for finite $\varepsilon$ in the blue region), which is visible in the currents.
Then, Pauli-blocking is negligible, no correlations can build up and the statistics appears Poissonian as with independent tunneling processes.
As the chemical potentials get closer to the end of the band gap (purple regime in panel (a)) and further states enter the transport window, the probability for transport blockade increases, 
such that the Fano factor drops and the statistics become sub-Poissonian (anti-bunched).
A similar phenomenon can be observed for the Fano factor in Fig.~\ref{fig:spectroscopy}(c) where we obtain $F=1$ inside the band gap as well.
Although there, the edge state is included into the transport window at any finite bias voltage, the transport takes place much deeper in the topological phase than in Fig.~\ref{fig:equalTemperature}. 
This leads to a very small edge state transmission which causes the same effect
that two electrons hardly ever compete for occupying the same edge state
as described above.

In contrast to Fig.~\ref{fig:spectroscopy}(c), we observe a slightly smaller Fano factor $F<1$ in Fig.~\ref{fig:equalTemperature} (b) inside the band gap where the edge state peak is situated between $\mu_L$ and $\mu_R$ (green area for finite $\varepsilon$, blue and green area for $\varepsilon=0$.
Here, the edge state transmission is much higher, which increases the probability for electrons to meet inside the chain and leads to sub-Poissonian transport as described above.

Eventually, Fig.~\ref{fig:equalTemperature} (c) shows that the heat currents entering the SSH system from either lead are for $\varepsilon=0$ 
always negative, meaning that heat is actually dissipated in both reservoirs. 
Therefore, we do not observe heat current flowing into the system at $\varepsilon=0$, and accordingly no cooling of the cold reservoir or power generation due to incoming heat currents. 
In contrast, for finite $\varepsilon$, both heat currents can become positive, which in presence of both a thermal and a temperature
gradient can be used for thermoelectric applications, see below.


\subsection{Thermoelectric generator and refrigerator}\label{sec:thermoelectric}

  \begin{figure}[b]
    \centering
     \setlength{\unitlength}{.1\linewidth}
	\includegraphics[width=1.\linewidth]{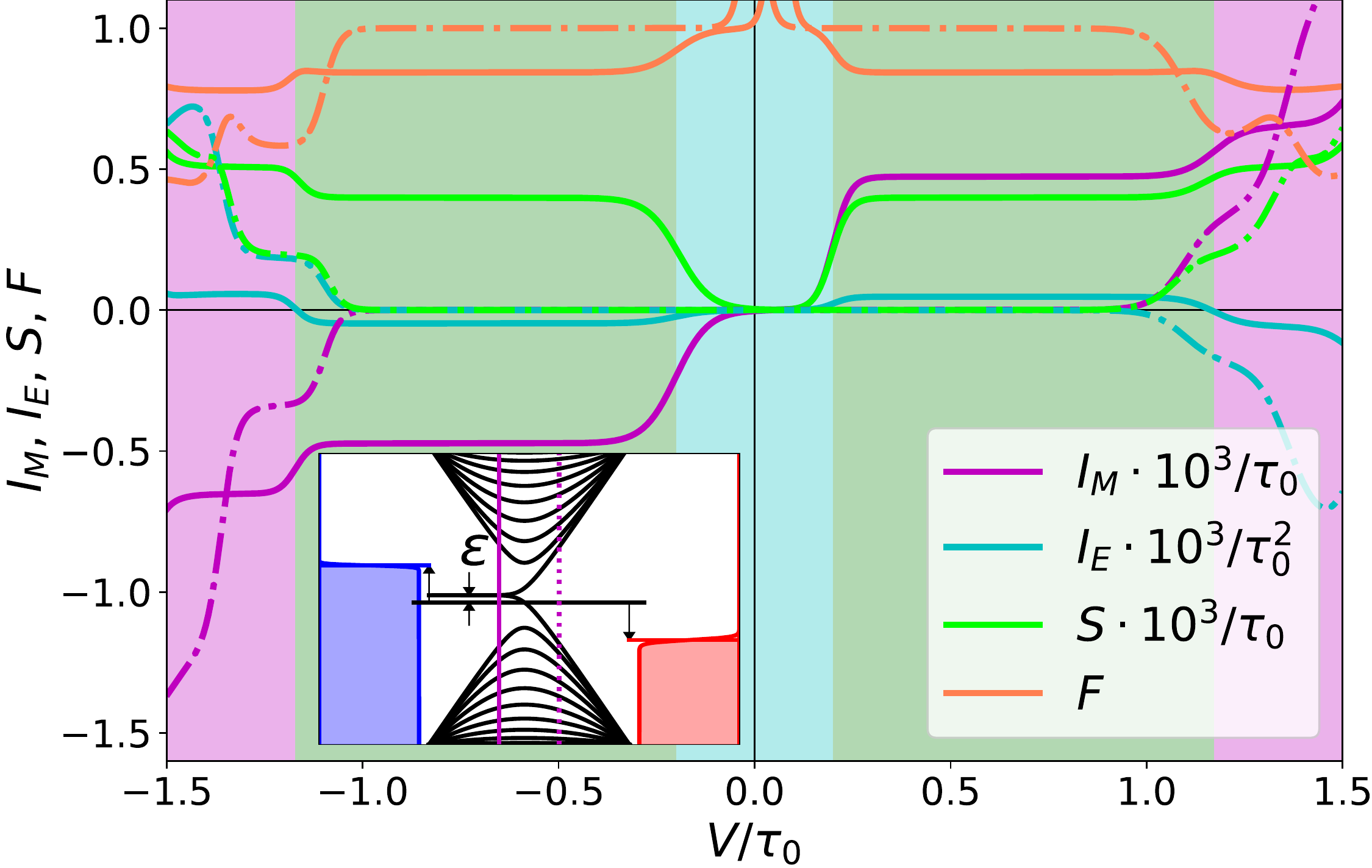}
   \caption{Particle current $I_M$, energy current $I_E$, noise $S$, and Fano factor $F$ for leads at finite temperatures $\beta_L = 100/\tau_0$ and $\beta_R = 50/\tau_0$, and in the presence of onsite potentials in the SSH chain $\varepsilon = 0.1 \tau_0$. The inset illustrates the transport window. Calculations were performed for $N=20$ sites ($10$ dimers) and reservoir couplings $\tau_L = \tau_R = 0.1\tau_0$. Dash-dotted lines refer to the trivial phase ($\delta\tau/\tau_0 = 0.3$), while solid lines refer to the topological phase ($\delta\tau/\tau_0 = -0.3$). }
    \label{fig:current}   
    \end{figure}

In this section, we aim to study the thermodynamical properties of the dimer chain.
We focus on the topological phase here and in particular on the edge states, as precisely there we find large efficiencies albeit at 
low total power output.
Correspondingly, we now consider both an electric bias $\mu_L-\mu_R=V \neq 0$ and a thermal bias $\beta_L>\beta_R$.
To operate the edge states in a heat engine, it is necessary to provide them with a finite energy:
This can be understood as we obtain $I_E \approx \varepsilon I_M$ for narrow transmission in the regime where only edge states participate in transport. 
Then, vanishing energy currents ($\varepsilon=0$) would by the second law always imply that the matter current will flow with the bias. 

Fig.~\ref{fig:current} depicts particle and energy current for this setup in its topological (solid) as well as trivial phase (dash-dotted).  
For bias voltages where only the edge states are included in the transport window (green (and turquoise, due to finite temperatures) region), the topological phase shows the discussed behavior necessary to obtain a heat engine.
%
%
In contrast, the trivial phase exhibits vanishing currents in this bias regime due to the absence 
of any states in the transport window.
Consistently, the Fano factor (dash-dotted orange) reaches $F\approx 1$ in this regime, 
accounting for Poissonian transport as it has been described before.

As soon as one of the chemical potentials reaches the bulk (purple regime), currents quickly reach values much higher than currents in the topological phase due to the higher transmission in the trivial phase.

\begin{figure}[ht]
    \centering
     \setlength{\unitlength}{.1\linewidth}
	\begin{picture}(10,10.65)
	\put(0.,5.3){\includegraphics[width=\linewidth]{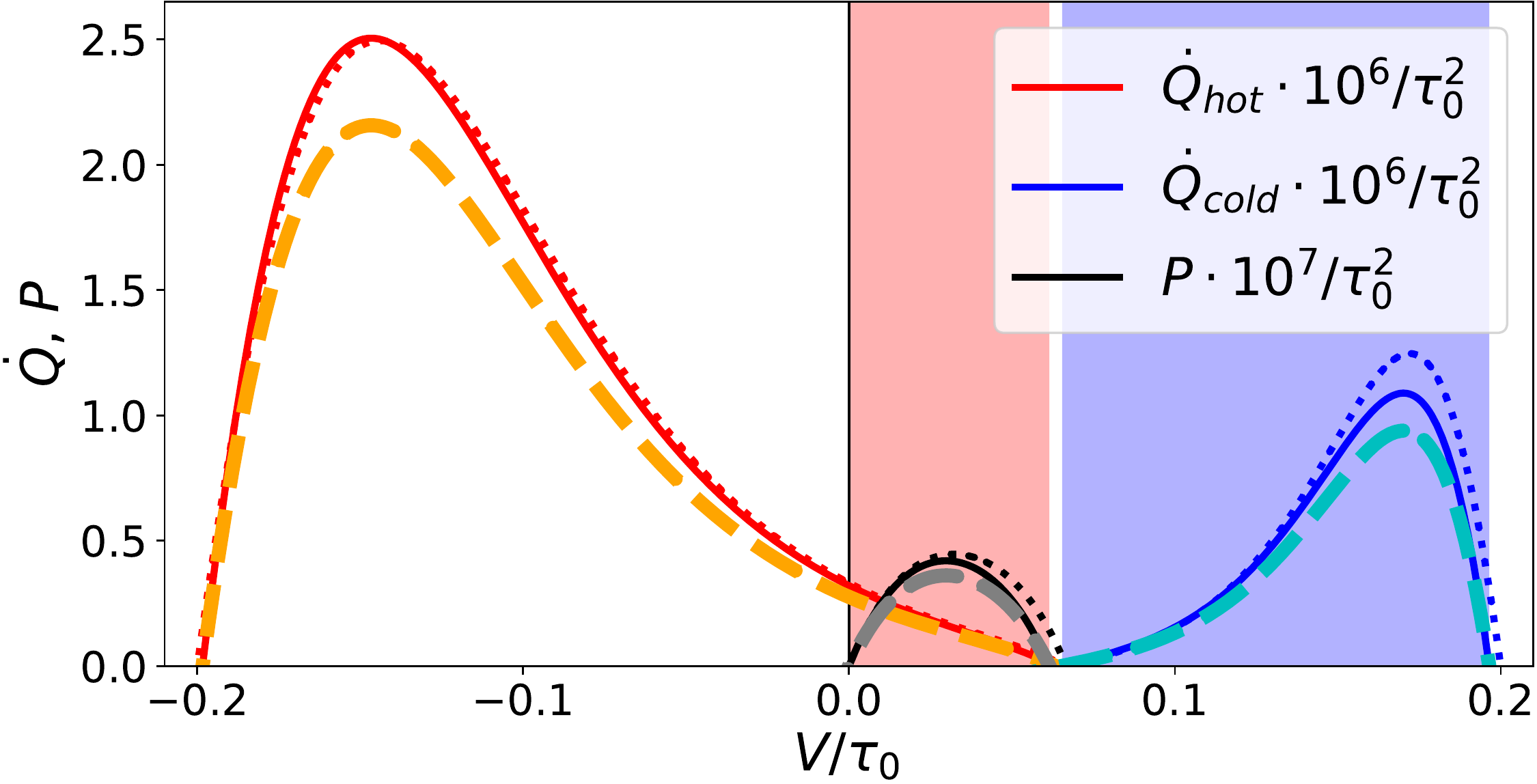}}
	\put(0.,0){\includegraphics[width=.66\linewidth]{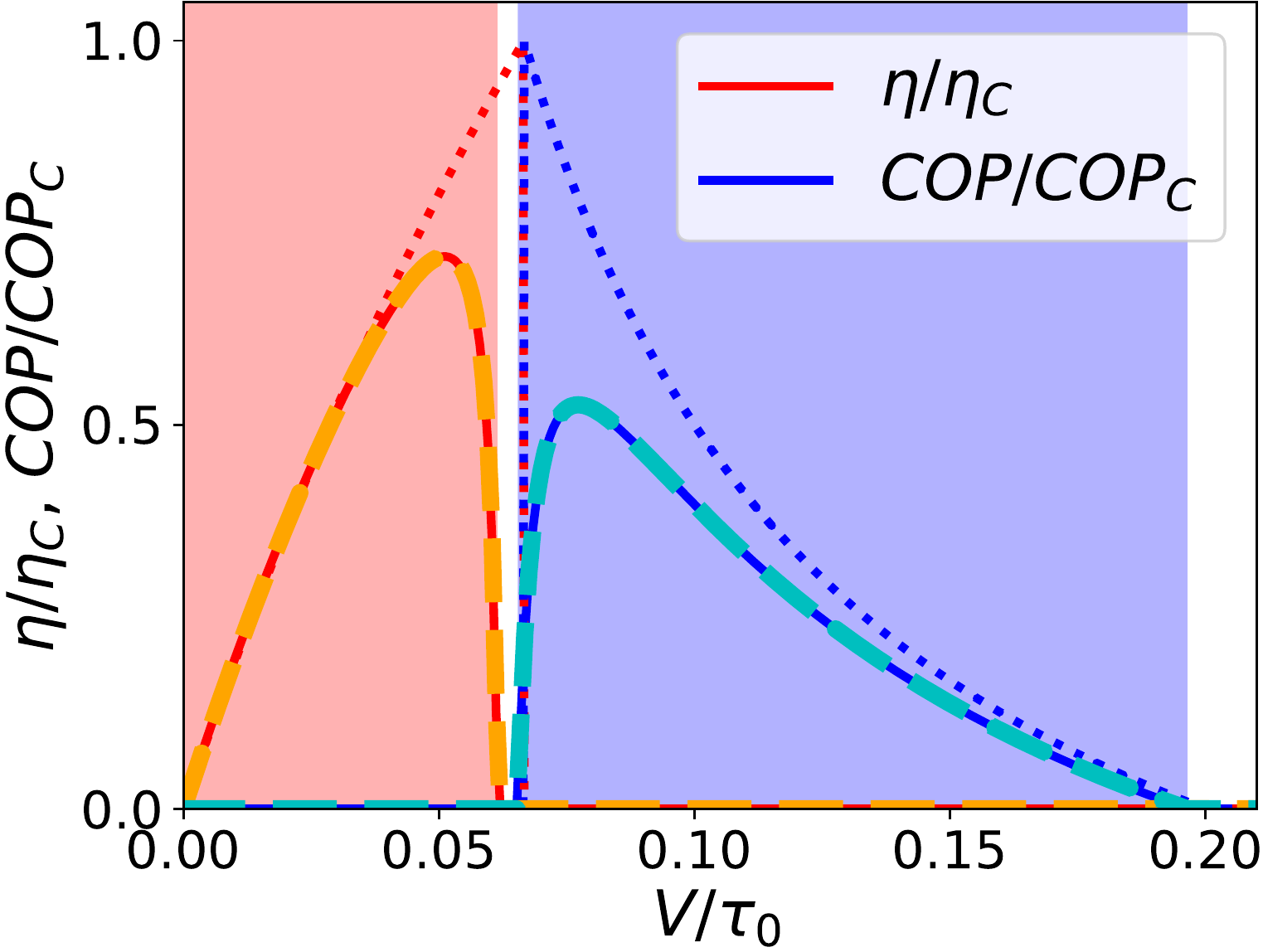}}
	\put(7.,0){\includegraphics[width=.33\linewidth]{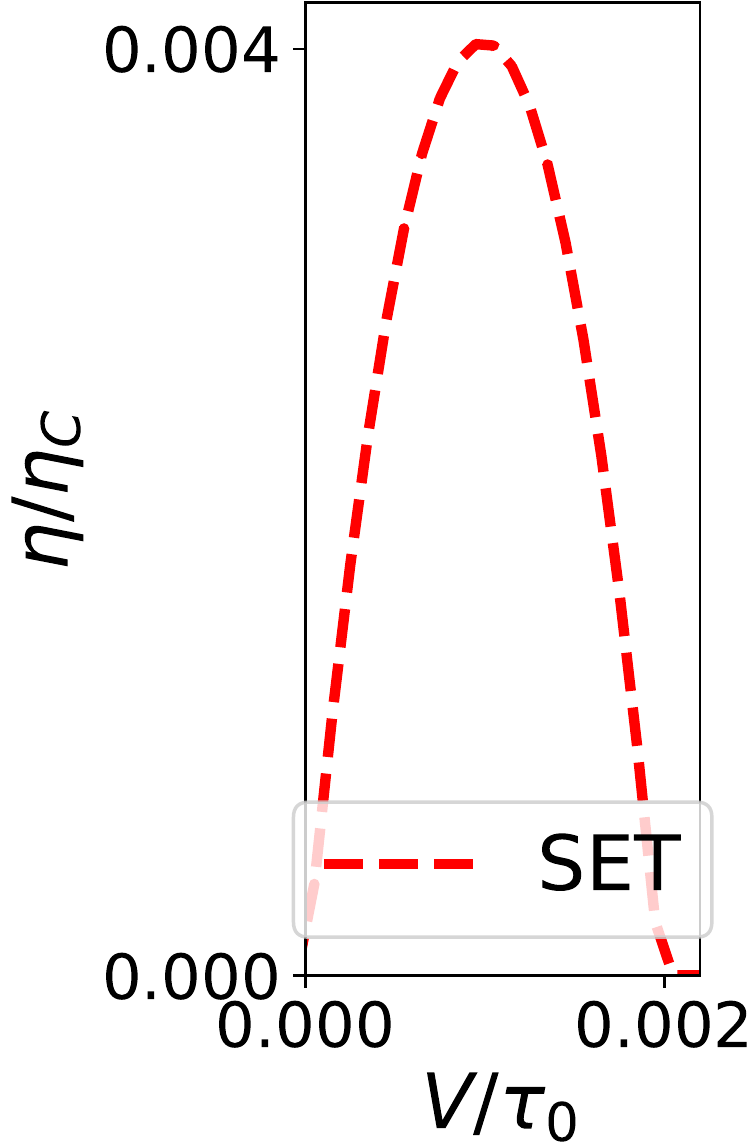}}
	\put(0.,10.6){\textbf{(a)}}
    \put(0.,5.2){\textbf{(b)}}
    \put(7.,5.2){\textbf{(c)}}
	\end{picture}
   \caption{Heat currents and thermodynamic performance of our setup in the topological phase with parameters as in Fig.~\ref{fig:current}. Dotted lines indicate the master equation solution, see Appendix~\ref{APP:toymodel}. 
Thick dashed lines illustrate the case of the same dimer chain in presence of a specific perturbation of strength $\chi = 0.2\tau_0$ as described in Sec.~\ref{sec:disorder}.
   Red and blue background colors mark heat engine and cooling regime, respectively. 
   \textbf{(a)} Heat currents and power. The master equation solution has been computed at $\gamma = 2.5\cdot 10 ^{-4}\tau_0$. \textbf{(b)} Efficiency of the heat engine and coefficient of performance of the refrigerator, both divided by their Carnot bounds. 
   \textbf{(c)} Numerical solution for the heat engine efficiency in case of the SSH chain being replaced by a single electron transistor (SET) with corresponding onsite energy $\varepsilon = 0.1\tau_0$. Coefficient of performance of the single electron transistor is not depicted as the present reservoir couplings $\tau_{L} = \tau_R = 0.1\tau_0$ lead to a situation where cooling of the cold reservoir is not achieved. 
   }
    \label{fig:heat}   
    \end{figure}
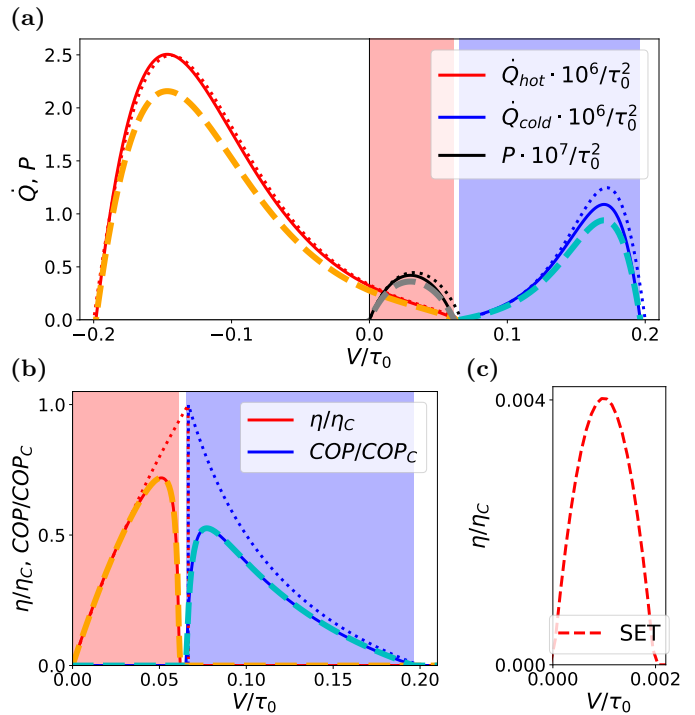

Results for heat currents $\dot{Q}$ and power $P$ are depicted in Fig.~\ref{fig:heat} (a). 
It shows that for small negative bias, heat will enter the system from the hot reservoir ($\dot{Q}_{\rm hot}>0$) and
simultaneously heat is further transferred into the cold reservoir ($\dot{Q}_{\rm cold}<0$, not depicted).
For more negative bias, both heat currents become negative, indicating that the bias is so strong that heat is transferred to both 
reservoirs.
Altogether, for negative bias voltages we always need to invest electric power ($P<0$, not depicted).
Now, for positive bias voltage, we can identify two regimes interesting for applications.

\textbf{Heat engine.} The red area is defined by $P>0$, and we see that there also $\dot{Q}_{\rm hot}$ is positive, which means that 
heat enters the chain from the hot reservoir. 
The generation of positive power is possible with a fraction of the heat coming from the hot reservoir.
The remaining fraction is passed on to the cold reservoir $\dot{Q}_{\rm cold}<0$.
In this regime (red), the chain works as a heat engine, effectively transporting electrons through the SSH chain against the bias.

\textbf{Refrigerator.} In the blue area, $\dot{Q}_{cold}$ is positive which indicates that heat is entering the system from the cold 
reservoir, cooling it like a refrigerator. 
For this, we have to invest chemical work ($P<0$), which is then dumped as heat into the hot reservoir.
In this regime, the heat is mainly dissipated into the hot reservoir while the particle current follows the voltage. 

Furthermore, Figs.~\ref{fig:heat} (a) and (b) compare the thermodynamic performance of our system to a
master equation solution (dotted lines) based on a minimal toy model which is described in App.~\ref{APP:toymodel}. 
The scenario illustrated by this toy model is a transport process governed exclusively by two edge states, 
localized to the left and to the right end of the chain, respectively, and neglecting any coherences between them.
An electron in the left edge state tunnels with a high tunneling rate $\Gamma$ into the left lead, and with a low tunneling rate $\gamma\ll\Gamma$ into the right lead. 
An analogous process takes place for the right edge state.

The efficiency of the thermoelectric generator in the red area and the coefficient of performance in the blue area are illustrated in Fig.~\ref{fig:heat} (b). 
Both values are divided by their Carnot bound so that they cannot exceed the value $1$. 
One can see that the dotted curves (toy model) and solid curves (exact calculations) are similar, which is a good benchmark 
for the validity of the reduced three-state model in this bias window.
We further see that the toy model actually reaches the maximum heating and cooling efficiencies, 
which the exact model calculations do not. This results from the level broadening taken into account by the
exact calculation.
For example, the SSH chain does not only allow transmission at $\varepsilon$, but also at energies slightly above or below $\varepsilon$, 
since the transmission peaks are not infinitely narrow.
Therefore, in the exact calculations, the energy current is not exactly proportional to the matter current (tight coupling), 
reducing all efficiencies.
This violation of tight-coupling also leads to the gap between heating and cooling function, which the master equation approach (which assumes tight coupling) does not have.
This breaking of the tight-coupling regime by finite coupling strengths is also known from other models beyond the weak system-reservoir coupling limit~\cite{restrepo2018a}.
Finally, the comparison with a single-electron transistor (SET, our setup with a single site $N=1$ only) shows that the thermoelectric performance
of the edge states exceeds that of a similarly strongly coupled single quantum dot.
In contrast to the edge states $\left| L\right>$, and $\left| R\right>$ which exhibit a rather narrow transmission peak because they are coupled only weakly to the reservoir on the other side of the chain, the transmission through a SET with the same reservoir coupling broadens to an extend that tight coupling is no longer present.


\subsection{Robustness against tunnel perturbations}\label{sec:disorder}
      
In order to test the robustness of the previously discussed nanothermal engines in presence of symmetry-preserving perturbations, we add a tunnel disorder term 
\begin{equation}\label{EQ:disorder}
H_{\rm D} = \sum_{j=1}^{N-1} \delta_j \left(\hat c_{j+1}^\dagger \hat  c_{j} + {\rm h.c.} \right)
\end{equation}
to our Hamiltonian $H$, where 
$\delta_j$ denotes a change of the hopping amplitudes within the SSH chain section.
For $\varepsilon=0$, this term still preserves the chiral symmetry, such that the relevant boundary modes should be expected robust with respect to such 
perturbations (again, finite $\varepsilon$ just induces a shift of the excitation energies).
To demonstrate this robustness explicitly, we specifically choose 
\begin{align}
\delta_j &= \left\{\begin{array}{lll}
        +\frac{\chi}{2} & : & 1\leq m < N/2\\
        0 & : &  m = N/2\\
        -\frac{\chi}{2} & : & N/2 < m < N
        \end{array}\right\}        	
\end{align}
where $\chi$ is a real number that quantifies the strength of the perturbation applied to the inter- and intra-dimer couplings inside the SSH chain consisting of $N$ sites.

The resulting thermoelectric performance of a dimer chain that is subject to perturbation of strength $\chi = 0.2\tau_0$ is shown in Fig.~\ref{fig:heat} (thick dashed lines). 
Even for a quite strong perturbation, we indeed observe a nearly perfect agreement of the efficiencies with the unperturbed chain, 
and the windows of both operational modes are hardly changed.
This would be different for disorder on the on-site energies (not shown), for which a stronger topological protection 
would be required.
We note that for periodically-driven SSH chains, a much more robust behaviour of boundary modes 
has been found~\cite{balabanov2017a}, which would be relevant for heat engines extracting mechanical instead of chemical work.
We discuss more general disorder types in Appendix~\ref{app:isolatedSSH}.


\subsection{Scalability and Implementation}\label{sec:scalability}

\begin{figure}[ht]
	\centering
	\includegraphics[width=1.\linewidth]{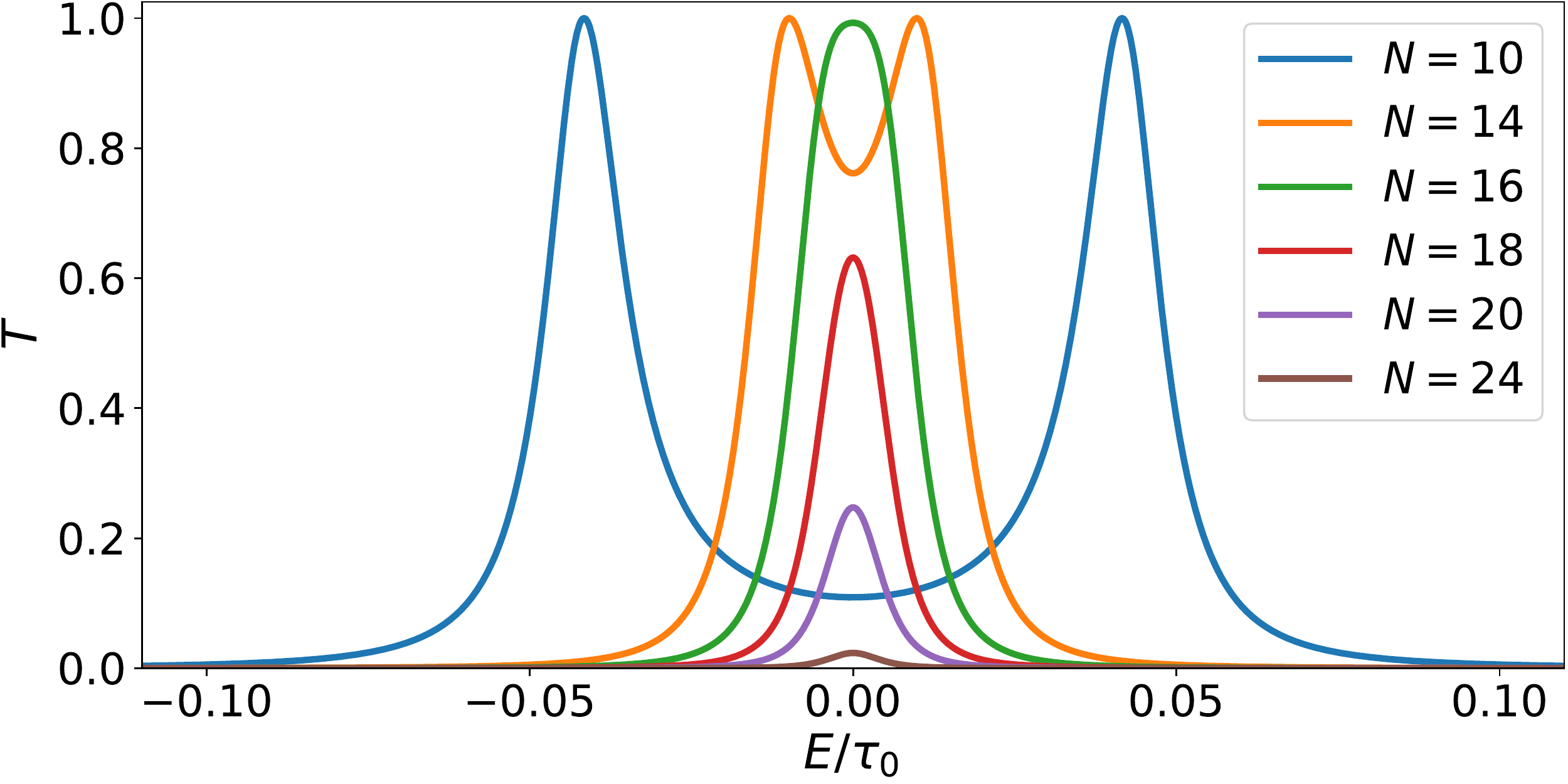}
	\caption{
	Transmission as a function of energy for different number of sites $N$ at $\delta\tau = -0.3\tau_0$ and reservoir coupling $\tau_L =\tau_R = 0.1\tau_0$ 
	and vanishing on-site energy $\varepsilon=0$.
	}
    \label{fig:scalability}   
	\end{figure}

Present experimental implementations have achieved quantum dot chains of up to 9 dots in
series ~\cite{zajac2016a}.
Absorbing one dot in the reservoirs, this would allow to create an SSH chain of four dimers, which is very
close to the visibility of topological effects.
In an ideal, scalable setup, the topological phase transition becomes more and more pronounced as the chain length is increased.
In such a setup, is important to consider that the precursor of the phase transition is dependent on $N$ as well, which is particularly
visible for smaller $N$.
For long chains, the edge states at the boundaries will be exponentially localized to the boundaries, such that their
contribution to the transmission deeply in the topological phase will be exponentially suppressed.
In these chains, one has to move the operational point extremely close to the topological phase transition
to obtain a non-vanishing transmission, whereas for shorter chains this is not so critical.
We have performed numerical calculations of the transmission $T(E)$ for fixed $\delta\tau$ for different $N$, see Fig.~\ref{fig:scalability}.
For very small $N$, the chosen value of $\delta\tau$ belongs to the normal phase, and one can correspondingly 
see two peaks of the innermost bulk states in the transmission with maximum at $T=1$ (blue and orange).
As $N$ is increased, the peaks -- still at maximal transmission -- collide as one runs through the precursor of the topological phase transition, 
where the innermost bulk states transform into edge states (green)
For even higher $N$, the transmission at $\omega=\epsilon$ then encodes the edge state transmission, which decays 
very fast as N is further increased (red, purple, brown).
Therefore, to maintain the same transmission with increasing $N$, it will be necessary to move $\delta\tau$ closer 
to the value where the phase transition takes place.
Given a scalable implementation of the SSH chain, choosing the optimal working point of the device will be a trade-off
between robustness and high efficiency (going deeper into the topological phase) and power output (maximizing transmission by approaching 
the boundary of the normal phase). 
While the trade-off between efficiency and power has been a long-standing problem in the design of quantum heat engines, 
it is interesting to see that efficiency and topological protection can be optimized simultaneously in our setup.


\section{Conclusions}
 \label{sec:conclusion}
 
We have analyzed electron, heat and energy transport through a dimer chain by considering the SSH Hamiltonian coupled to two fermionic reservoirs subject to a finite DC bias voltage. 
By means of Green's functions techniques we solved the non-interacting many body Hamiltonian and investigated the 
thermoelectrical properties arising in a non-equilibrium setup.
Specifically, we have analyzed signatures of the underlying topological phase transition in a nonequilibrium scenario.

In the limit where the SSH chain is weakly coupled to the reservoirs, a transport spectroscopy picture applies and the excitation spectrum of the
SSH chain directly maps to steps in the electronic current, provided the temperature of the reservoirs is sufficiently low.
Although the edge states are only weakly coupled to one reservoir due to their localization to one end of the SSH chain at finite bias voltage 
(e.g., $\left| L\right>$ couples only very weakly to the right reservoir), their signature is visible for finite chain lengths as a small first plateau in the currents.
Thereby, topological and trivial phases can be clearly distinguished by electronic transport spectroscopy. 
Moreover, for finite-size chains and sufficiently low temperatures the individual excitation energies can be directly inferred from the electronic current.
Noise and Fano factor also mirror these observations:
%
Pauli blocking leads to an anti-bunching in the electronic transport statistics, 
with Fano factors below one for regions with non-negligible currents.

Regarding the stationary SSH occupation, we have investigated the scenario of a highly asymmetric but weak reservoir coupling strengths on the microscopic (Hamiltonian) level in combination with 
specific bias configurations, where we found that one of the edge states can be preferentially prepared as the stationary state of the dissipative non-equilibrium setup.
There, we could verify that the onset of the edge state formation at the topological phase transition becomes
sharper with increasing SSH chain length.

To implement heat engines, one could also think of utilizing bulk states, which
are less robust against perturbations.
In addition, they are for long chains energetically close to other bulk states and 
can hardly be put individually in a transport window. 
The resulting device would have a poor thermoelectric performance as energy and matter currents are not proportional.
Therefore, we focused our discussion on the thermoelectric performance of the edge states in the topological phase.
Efficiencies and coefficient of performance of heat engine and cooling operational modes can be significantly increased 
by achieving energy and matter currents to be tightly coupled $I_E = \varepsilon I_M$.
With the SSH model, the edge states in the topological phase are well-separated from the rest of the spectrum and can be selectively excited by
applying a small bias.
However, to put them to labor in a heat engine with symmetric bias assumptions $\mu_L=-\mu_R$, it was necessary that they 
have a finite energy $\varepsilon$.
Even for moderate symmetric or unsymmetric chain-reservoir coupling strengths, we observed that the edge states are effectively weakly coupled to one of the reservoirs and have a very sharp transmission peak, thereby realizing the tight-coupling regime with high precision.
For them, the SSH chain acts like a precise energy filter that implements a perfect tight-coupling scenario
with large heat engine efficiencies and cooling coefficients of performance.
These significantly exceed those of a single quantum dot with comparable coupling strengths, at which it is not in the
tight-coupling regime.

In this work, we have demonstrated that the topologically protected edge states can serve as the working medium of a robust quantum heat engine.
Here, the position of the edge state energies $\varepsilon$ (and the resulting tight-coupling relation $I_E = \varepsilon I_M$)
is protected by topology with an exponential accuracy in the length of the total chain.
Nevertheless, the total power produced may still sensitively depend on the optimal working point $\delta\tau$.
We hope that our discussion of the benefits of topology in one-dimensional atomic arrays paves the way for 
studies on the design of quantum engines in more complicated systems with higher dimensions and nontrivial topology.

\acknowledgments{
Our study of SSH chains has been initiated by Prof. Tobias Brandes to whom we devote this work. 
We warmly acknowledge him for enlightening discussions. 
The authors have profited from discussions with J. Cerrillo, \'{A}. G\'{o}mez-Le\'{o}n, and S. Restrepo.

Furthermore, financial support by the DFG (SFB 910, GRK 1558, BR 1528/9-1) is gratefully acknowledged.
}


\bibliography{literature}


\appendix

\begin{widetext}


\section{The isolated SSH chain}\label{app:isolatedSSH}

The isolated SSH chain consists of a one-dimensional dimer chain where inter- and intra-dimer couplings alternate, as described by the tight-binding Hamiltonian in Eq.~(\ref{EQ:ham_ssh}).
The effect of finite $\varepsilon$ is a shift of the electronic excitation energies, it will be neglected in our introductory
discussion of the model.
The topological phase transition caused by the alternating hopping amplitude $t_\pm\equiv\tau_0 \pm \delta\tau$ can be observed in the single-particle excitation spectrum depicted in the inset of Fig.~\ref{fig:spectrum} (a). 

There, we observe two regions where the energies are rather dense for all values of $\delta \tau$, which in the continuum limit ($N\to\infty$)
become bands.
Most importantly, we find that in the middle of the gap between the bands, two energies merge for $\delta\tau<0$ at energy $E=\varepsilon$. 
The wave functions of these {\em midgap modes} are depicted in the main panel. 
We observe that they are strongly localized close to the boundaries, which justifies to denote them as boundary or edge modes. 
For example, on the left part of the SSH chain, their  wave function reads~\cite{Asboth2016} approximately $\psi(j )\propto (1- (-1)^j )(t_-/t_+)^{(j/2-1/2) }\psi_0$, where 
$\psi_0 \in \mathbbm C$ accounts for the normalization. 
Thus, the wave function exhibits an exponential decay along the chain for odd sites and vanishes strictly on all even sites, which is also visible in Fig.~\ref{fig:spectrum} (a).
 
 \begin{figure}[ht]
    \centering
     \setlength{\unitlength}{.1\linewidth}
	\begin{picture}(10,4.1)
	\put(0.,0.){\includegraphics[width=.49\linewidth]{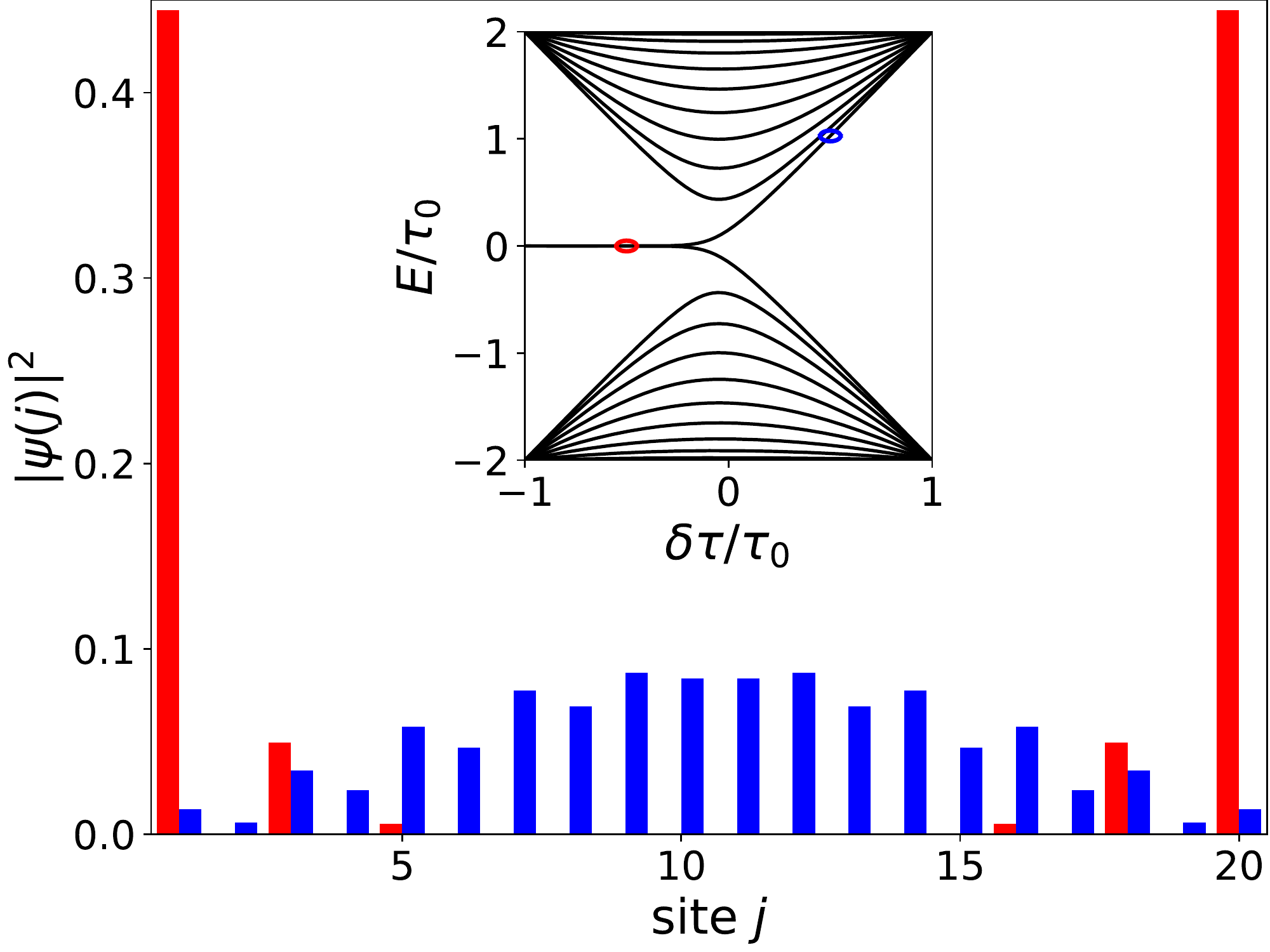}}
	\put(5.,0.){\includegraphics[width=.49\linewidth]{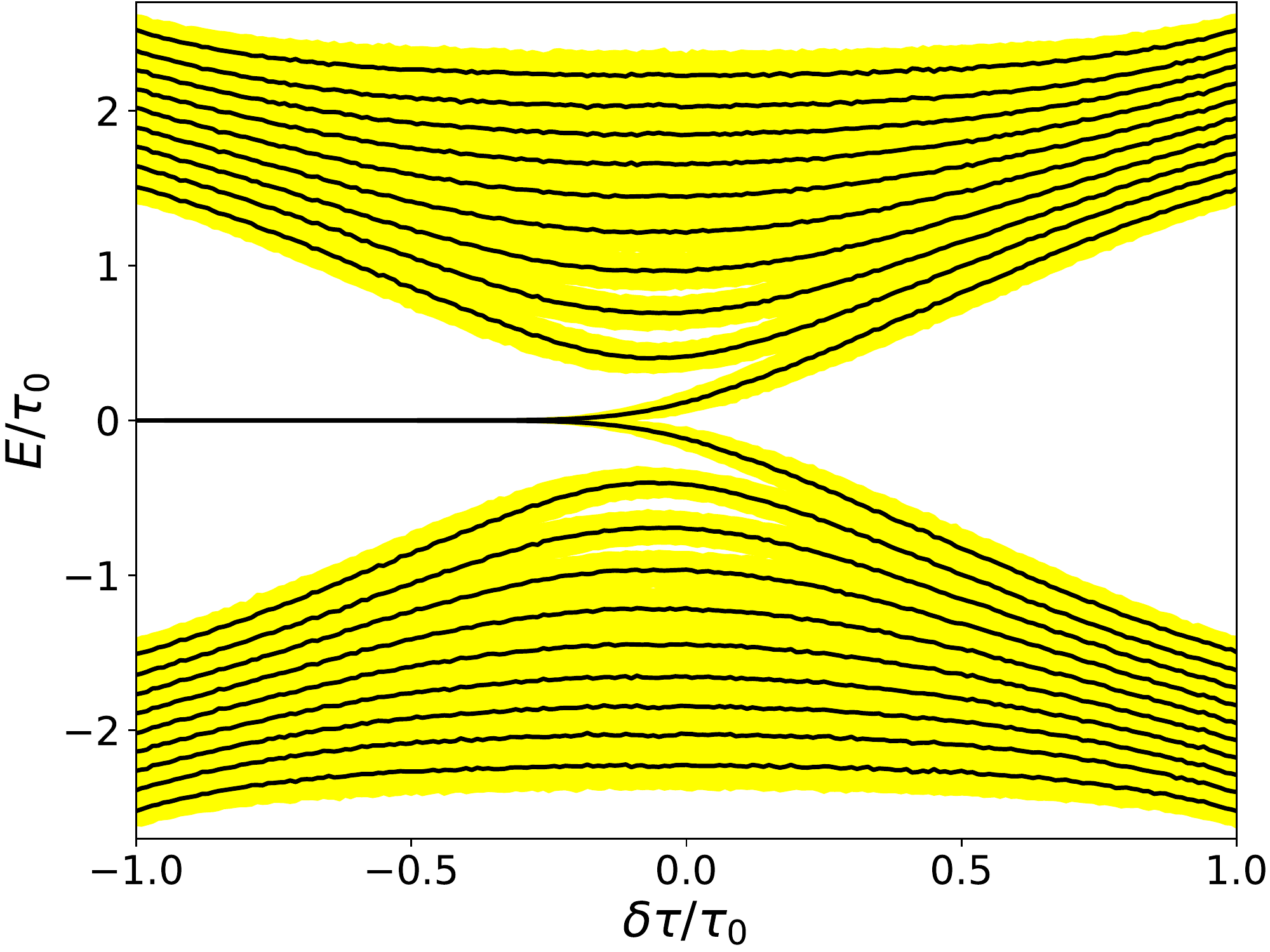}}
	\put(0.,3.5){\textbf{(a)}}
	\put(5.1,3.5){\textbf{(b)}}
	\end{picture}     	
\caption{\label{fig:spectrum} 
Spectral properties of an SSH chain with $N=20$ sites and $\varepsilon=0$.
\textbf{(a)}
Spacial distribution $\abs{\Psi(j)}^2 \equiv \abs{\bra{\Psi} c_j^\dagger c_j \ket{\Psi}}^2$ of two selected eigenstates $\ket{\Psi}$ 
of the central SSH Hamiltonian~(\ref{EQ:ham_ssh}), taken at 
$\delta \tau = -0.5\tau_0$ (red) and $\delta \tau = 0.5\tau_0$ (blue). 
The inset displays the energy spectrum
with red and blue circles indicating the energy states $\ket{\Psi}$ shown in the main plot.
\textbf{(b)}
Average of 1000 spectra subject to tunnel disorder with $\delta_j$ uniformly distributed in the interval $[-\tau_0/2,+\tau_0/2]$, 
compare Eq.~(\ref{EQ:disorder}).
The yellow region in the background represents the mean plus/minus one standard deviation, and one can see that only the edge states 
are completely robust to these perturbations.
}
\end{figure}
Furthermore, the spectrum is symmetric with respect to $E=\varepsilon=0$, which is a strict consequence 
of the chiral symmetry present in this system. 
For $\varepsilon=0$, the chiral symmetry relation reads 
 \begin{equation}
 \Sigma H_{\rm SSH}\Sigma = -H_{\rm SSH}\,,
 \end{equation}
where the chiral symmetry operator is given by~\cite{Asboth2016} $\Sigma= \exp\left\{\ii \frac{\pi}{2} \sum_j [1-(-1)^j] \hat c_{j}^\dagger \hat  c_{j}\right\}$.
Finite $\varepsilon$ just shift the excitation spectrum.

The two continua of states are related to the two bands of a corresponding infinite system $N \rightarrow \infty$.
One can show that the infinite system is characterized by a topological invariant $W$, which assumes the values $W=0$ and $W=1$ 
for $\delta \tau >0$ and $\delta\tau <0$, respectively. 
Thus, there is a one-to-one relation of existing pairs of boundary modes and the topological invariant.
This specific example is a manifestation of the general bulk-boundary correspondence ubiquitously appearing in 
the field of topological band structures~\cite{Gurarie2011}. 
 
Deeply in the topological phase, the chiral symmetry protects the two  boundary modes from energetically splitting and leaving energy $E=\varepsilon$. 
More precisely, the splitting vanishes for $\delta\tau<0$ exponentially with the chain length~\cite{Asboth2016}
$\Delta E  \propto e^{-N}$.
This property is robust with respect to perturbations that obey these symmetries, which justifies to call these modes topologically protected.
Moreover, due to the inversion symmetry in the finite-size SSH model, the eigenstates of~(\ref{EQ:ham_ssh}) are either even
or odd upon inversion with respect to the middle of the chain. 
Correspondingly, also the two midgap states can be written as a symmetric and an antisymmetric superposition 
$\left|\pm\right> = \frac{1}{\sqrt{2}} \left( \left|L\right> \pm \left|R\right> \right)$ of states  $\left|L\right>$ and  $\left|R\right>$ 
which are localized at the left and right ends of the SSH chain, respectively.

Following the level flow of one of the edge states into the topologically-trivial phase $\delta \tau >0$, we find that they 
approach the remaining energy states in the bulk, as can be observed in the inset of Fig.~\ref{fig:spectrum} (b). 
The other states in the dense energy region extend in a similar fashion over the bulk (not shown). 

Finally, we mention that at the point of the topological phase transition $\delta\tau=0$, the SSH model falls back onto the analytically solvable homogeneous chain, 
which we discuss in Appendix~\ref{app:spectralCouplingDensities}.

The discussed chiral symmetry even holds in the presence of tunnel disorder.
We demonstrate this in Fig.~\ref{fig:spectrum} (b), where we see that when adding a random tunnel disorder 
Hamiltonian~(\ref{EQ:disorder}) with $\delta_j$ uniformly distributed in the interval $[-\tau_0/2, +\tau_0/2]$, 
the bulk states may be shifted but the boundary modes are pinned to value $\varepsilon$.
Accordingly, we see after averaging that the bulk state energies aquire a finite variance (yellow background), 
whereas the variance of the topologically protected boundary modes vanishes.


\section{Green's function calculation}\label{APP:greens_function}

The Green's function applied in our calculations can easily be obtained if we split the full Hamiltonian $H$ into two parts $H=H_0+H_1$, where the Green's function
of $H_0$ should be known.
Specifically, we consider an infinite tight-binding chain with nearest neighbor coupling $\tau_0$
\begin{equation}
H_0 =  \sum_{j=-\infty}^{\infty} \tau_0 \left( \hat c_{j+1}^\dagger \hat  c_{j} + {\rm h.c.}\right)\,.
\end{equation}
Consequently, the remaining part must include all additional terms
\begin{equation}
H_1 = \left( (\tau_{\rm L} - \tau_0) \hat c_{1}^\dagger \hat  c_{0} +(\tau_{\rm R}- \tau_0) \hat c_{N}^\dagger \hat  c_{N+1}+ {\rm h.c.}\right) 
+ \sum_{j=1}^{N}\left[\varepsilon \cdot \hat c_{j}^\dagger \hat  c_{j}- (-1)^j\delta\tau \left( \hat c_{j+1}^\dagger \hat  c_{j} + {\rm h.c.}\right)\right]\,.
\end{equation}
The advantage of this splitting is that the free system can be diagonalized by Fourier transform 
\begin{equation}
c_n = \frac{1}{\sqrt{2\pi}} \int_{-\pi}^{+\pi} c(\kappa) e^{-\ii n \kappa} d\kappa\,,\qquad
c(\kappa) = \frac{1}{\sqrt{2\pi}} \sum_n c_n e^{+\ii n \kappa}\,,
\end{equation}
and the free Hamiltonian then becomes $H_0 = \int_{-\pi}^{+\pi} \varepsilon_\kappa c^\dagger(\kappa) c(\kappa) d\kappa$
with $\varepsilon_\kappa = 2\tau_0 \cos(\kappa)$.
Due to its non-interacting structure, it suffices to consider the eigenvectors $\left| \kappa \right) = c^\dagger(\kappa) \left|0\right)$ and eigenvalues $\varepsilon_\kappa$ 
of the single-particle physics (we denote the single-particle sectors by $\mathcal{H}_0$ and $\mathcal{H}_1$), and the Green's function is given by its spectral representation 
\begin{equation}
G_0 (E) = \left[E-\mathcal{H}_0\right]^{-1} 
= \int_{-\pi}^{+\pi} \frac{  \left| \kappa \right) \left( \kappa \right| } {E-\varepsilon_\kappa} d\kappa\,.
\end{equation}
This expression is not well defined yet, since the denominator might contain zeros. 
This can be resolved by introducing an imaginary shift for retarded and advanced Green's functions as we did in Eq.~(\ref{eq:gRetAdv}).
For example, the retarded Green's function becomes in position space 
\begin{equation}
G_0^r(l,m;E) = \left(0\right| c_l G_0^r(E) c_m^\dagger \left|0\right) = \lim_{\delta\to 0^+} \frac{1}{2\pi} \int_{-\pi}^{+\pi} \frac{e^{-\ii (l-m) \kappa}}{E-2\tau_0 \cos(k)+\ii \delta} d\kappa\,.
\end{equation}
Closing the contour with the appropriate case distinctions for $l,m$, we see that the integral contributions along $k=\pm\pi + \ii \sigma$ cancel, 
such that we can evaluate the integral with the residue theorem.
Eventually, one obtains a matrix in position space whose element ($l$, $m$) reads for $-2\tau_0 \le E \le +2\tau_0$
\begin{equation}
G_0^{r} (l,m; E)= \frac{-\ii}{\sqrt{4 \tau_0^2-E^2}}\cdot \left(\frac{E}{2\tau_0}-\ii\sqrt{1-\frac{E^2}{4\tau_0^2}}\right)^{|l-m|}\,,
\end{equation}
and the complex conjugate for the advanced Green's function.
Applying the defining equations of the Green's function, one can express the Green's function of the full system $G(E)$ in terms of the known Green's function of the $\mathcal{H}_0$ 
system due to (now suppressing for simplicity the labels of retarded and advanced Green's functions) 
\begin{equation}
G(E) = (E-\mathcal{H}_0-\mathcal{H}_1)^{-1}  
= \left[(E-\mathcal{H}_0)(\f{1}-(E-\mathcal{H}_0)^{-1}\mathcal{H}_1)\right]^{-1}
= (\f{1}-G_0(E)\mathcal{H}_1)^{-1}G_0(E)\,.
	 \label{eq:gDyson}
\end{equation}
Now, due to the local structure of $\mathcal{H}_1$, the matrix that needs to be inverted has only dimension $N+2$ for an SSH chain consisting of $N$ sites. 


\section{Spectral coupling density of the leads}\label{app:spectralCouplingDensities}

The calculation of the spectral coupling density requires the diagonalization of the reservoirs, which can be done in an exact fashion for the
considered system.
Therefore, we consider here the form of the reservoir Hamiltonians 
\bea\label{EQ:ham1}
H_{\rm lead} = \epsilon \sum_{i=1}^N \hat c_i^\dagger \hat c_i + \tau_0 \sum_{i=1}^{N-1} \hat c_i \hat c_{i+1}^\dagger + \tau_0 \sum_{i=1}^{N-1} \hat c_{i+1} \hat c_i^\dagger\,,
\eea
which describes a chain with $N$ sites, homogeneous next-neighbour hopping amplitudes $\tau_0$ and on-site energies $\epsilon$.
Without loss of generality, we consider $\tau_0$ as real-valued here. 
If it was not real from the beginning, we could rotate the annihilation and creation operators by a phase that removes the
complex phase from the tunneling amplitude.

Since the tunneling amplitudes are uniform along the lead, we can by using a specific (unitary) Bogoliubov transformation
\bea\label{EQ:bogoliubov}
\hat c_i = \sum_{k=1}^N u_{ik} \hat d_k\,,\qquad
\hat c_i^\dagger = \sum_{k=1}^N u_{ik}^* \hat d_k^\dagger\,,
\eea
where
\bea
u_{ik} = \sin\frac{\pi i k}{N+1}
\eea
fully diagonalize the lead Hamiltonian
\bea\label{EQ:lead_eigenvalues}
H_{\rm lead} = \sum_k \Omega_k \hat d_k^\dagger \hat d_k\,,\qquad
\Omega_k = \epsilon - 2 \tau_0 \cos\frac{\pi k}{N+1}\,.
\eea
We see that the $\Omega_k$ are centered around $\epsilon$ in the interval $[\epsilon-2\tau_0, \epsilon+2 \tau_0]$, and one
can compute the corresponding spectral density in the continuum limit $N\to\infty$.

If we do now couple the first site of the lead chain to some other system 
(in the main article, the SSH chain) via
\bea
H_I = \tau \hat c_1 \hat c^\dagger + \tau^* \hat c \hat c_1^\dagger\,,
\eea
where $\tau$ denotes the hopping amplitude to the site described by annihilation operator $\hat c$, 
we can also insert the very same Bogoliubov transformation, yielding
\bea
H_I &=& \tau \sum_{k=1}^N u_{1k} \hat d_k \hat c^\dagger + \tau^* \hat c \sum_{k=1}^N u_{1k}^* \hat d_k^\dagger\nn
&=& \tau \sum_k \sqrt{\frac{2}{N+1}} \sin \frac{\pi k}{N+1} \hat d_k \hat c^\dagger
+ \tau^* \sum_k \sqrt{\frac{2}{N+1}} \sin \frac{\pi k}{N+1} \hat c \hat d_k^\dagger\,.
\eea
This also defines the spectral coupling density of the new model, respectively
\bea
\Gamma(E) &=& 2\pi \sum_k \abs{t_k}^2 \delta(E-\Omega_k)
= 2\pi \sum_k \frac{2\abs{T_0}^2}{N+1} \sin^2 \frac{\pi k}{N+1} \delta\left(E-\Omega_k\right)\,.
\eea
To map this into a continuous distribution as $N\to\infty$, we can integrate along the 
interval $[(\Omega_k+\Omega_{k-1})/2, (\Omega_k+\Omega_{k+1})/2]$ containing exactly one
eigenvalue -- cf. Eq.~(\ref{EQ:lead_eigenvalues}) -- which collapses all but one
of the terms, yielding
\bea
\Gamma(\Omega_k) &=& \frac{4\pi \abs{t_k}^2}{\Omega_{k+1} - \Omega_{k-1}} = \frac{2\pi \abs{t_k}^2}{\frac{\Delta\Omega}{\Delta k}}\,.
\eea
In the infinite-size reservoir limit $N\to\infty$, we have
\bea
\frac{\Delta\Omega}{\Delta k} \to \frac{d\Omega}{dk} = \frac{2\pi \tau_0}{N+1} \sin\frac{\pi k}{N+1}\,.
\eea
To represent this as a function of $\Omega$, we have to solve the lead eigenvalues~(\ref{EQ:lead_eigenvalues}) for $k$
\bea
k = \frac{N+1}{\pi} \arccos\frac{\epsilon-\Omega}{2\tau_0}\,,
\eea
and eventually need to collect the leading orders for $N\to\infty$.
For our model at hand, this implies
\bea\label{EQ:specdenschain}
\Gamma(E) &=& \frac{\abs{\tau}^2}{\tau_0^2} \sqrt{4 \tau_0^2 - (E-\epsilon)^2} \Theta(E+2\tau_0-\epsilon)
\cdot\Theta(2\tau_0+\epsilon-E)\,.
\eea
This describes a semicircle spectral coupling density, which we have used in the Green's function calculations in the main article, 
where $\tau$ is then replaced by $\tau_L$ and $\tau_R$, respectively, and $\epsilon=0$.


\section{Minimal toy model}\label{APP:toymodel}

Guided by the intuition that we can reach a parameter regime where only the vaccuum state and the two edge states 
(left- and right-dominated) participate in transport, we can for this regime set up a minimal toy model based on a simple rate
equation picture.
We expect this toy model to hold when the coupling between the SSH chain and the leads is negligible compared to all other
parameters and when edge state coherences can be neglected.
The corresponding transition rates can be phenomenologically constructed from using the detailed balance principle, the Fermi functions in the
leads, and the observation that the left-dominated edge state is coupled strongly to the left lead and weakly to the right and
vice versa for the right edge state.
The corresponding generalized rate equation for the probabilities of finding the SSH chain in the vacuum ($P_0$), left edge state
($P_L$), or right edge state ($P_R$) reads
\bea
\left(\begin{array}{c}
\dot{P}_0\\
\dot{P}_L\\
\dot{P}_R
\end{array}\right)
=
\left(\begin{array}{ccc}
-(\gamma+\Gamma) (f_L+f_R) 
& \Gamma (1-f_L) e^{-\ii\chi} e^{-\ii\varepsilon\xi} + \gamma (1-f_R) 
& \gamma (1-f_L) e^{-\ii\chi} e^{-\ii\varepsilon\xi} + \Gamma (1-f_R)\\
\Gamma f_L e^{+\ii\chi} e^{+\ii\varepsilon\xi} + \gamma f_R
& -\Gamma (1-f_L) - \gamma (1-f_R) 
& 0\\
\gamma f_L e^{+\ii\chi} e^{+\ii\varepsilon\xi}  + \Gamma f_R 
& 0
& -\gamma (1-f_L) - \Gamma (1-f_R)
\end{array}\right)
\left(\begin{array}{c}
P_0\\
P_L\\
P_R
\end{array}\right)\,.
\eea
Here, $\gamma$ is the small tunneling rate of the left edge state to the right lead and of the right edge state to the left lead, 
and $\Gamma$ is the large tunneling rate holding for the left edge state to the left lead and the right edge state to the right lead.
Furthermore, $f_\alpha=[e^{\beta_\alpha(\varepsilon-\mu_\alpha)}+1]^{-1}$ denotes the Fermi function of lead $\alpha$, and $\varepsilon$ 
the energy of the edge state.
The counting fields $\chi$ and $\xi$ for electrons jumping from the left lead onto the SSH chain and for the energy transferred from the left lead onto the SSH chain, respectively, can now be used to evaluate e.g. mean currents and noise within the validity range of the toy model.
Since there is only the single allowed transition frequency $\varepsilon$, within the three-state model energy and matter currents obey
tight coupling, i.e., the energy current is directly linked to the matter current via $I_E = \varepsilon I_M$.

In particular, we get when $\gamma \ll \Gamma$ the simplified expressions for matter current and noise, respectively
\bea
I_M = \gamma \frac{2-f_L-f_R}{1-f_L f_R} (f_L-f_R)\,,\qquad
S = \gamma \frac{2-f_L-f_R}{1-f_L f_R} (f_L+f_R-2 f_L f_R)\,.
\eea
From the matter current and the derived energy current $I_E = \varepsilon I_M$, we can compute power and all heat currents, 
which eventually leads to the dotted curves in Fig.~\ref{fig:heat}.
We also see that although the dependence of the current on the Fermi functions is different from that of a conventional single electron transistor (SET), where
in the master equation limit we have $I_M^{\rm SET} = \gamma (f_L-f_R)$, due to the tight-coupling between energy and matter currents in this limit, we would 
obtain precisely the same heat engine efficiency or coefficient of performance as for the SET.
However, in practical applications the efficiency at maximum power is more relevant, and we find numerically that our toy model is
slightly more advantageous than an SET. 
Furthermore, the noise from the toy model is always larger than the noise of a correspondingly asymmetric SET, such that the increased
noise level could in principle be used to identify multiple degenerate states participating in transport.

\end{widetext}

\end{document}